\documentclass[aip,pop,preprint]{revtex4-2}
\usepackage{amsmath}
\usepackage{amssymb}
\usepackage{graphicx}
\usepackage{hyperref}
\graphicspath{{figures/}} 
\usepackage[usenames, dvipsnames]{color}
\usepackage[latin1]{inputenc}
\usepackage{tikz}
\usetikzlibrary{trees}
\usetikzlibrary{decorations.pathmorphing}
\usetikzlibrary{decorations.markings}
\usepackage[compat=1.1.0]{tikz-feynman}
\usepackage{soul}

\usepackage{todonotes}
\newcounter{todocounter}

\newcommand{\vecu}{\mathbf{u}}

\newcommand{\real}{\operatorname{\mathbb{R}e}}

\newcommand{\ux}{\hat{u}_x}
\newcommand{\uy}{\hat{u}_y}
\newcommand{\uz}{\hat{u}_z}

\begin{document}
\author{B.~Tripathi$^1$}         
\email{btripathi@wisc.edu}
\author{P.W.~Terry$^1$}          
\author{A.E.~Fraser$^{2,3,4}$}   
\author{E.G.~Zweibel$^{1,5}$}    
\author{M.J.~Pueschel$^{6,7}$}   
\affiliation{
$^1$Department of Physics, University of Wisconsin-Madison, Madison, Wisconsin 53706, USA\\
$^2$Department of Applied Mathematics, University of Colorado, Boulder, Colorado 80309, USA\\
$^3$Department of Astrophysical and Planetary Sciences, University of Colorado, Boulder, Colorado 80309, USA\\
$^4$Laboratory for Atmospheric and Space Physics,, University of Colorado, Boulder, Colorado 80309, USA\\
$^5$Department of Astronomy, University of Wisconsin-Madison, Madison, Wisconsin 53706, USA\\
$^6$Dutch Institute for Fundamental Energy Research, 5612 AJ Eindhoven, The Netherlands\\
$^7$Eindhoven University of Technology, 5600 MB Eindhoven, The Netherlands\\
}

\title{Three-dimensional shear-flow instability saturation via stable modes}

\today

\begin{abstract}

Turbulence in three dimensions ($3$D) supports vortex stretching that has long been known to accomplish energy transfer to small scales. Moreover, net energy transfer from large-scale, forced, unstable flow-gradients to smaller scales is achieved by gradient-flattening instability. Despite such enforcement of energy transfer to small scales, it is shown here not only that the shear-flow-instability-supplied $3$D-fluctuation energy is largely inverse-transferred from the fluctuation to the mean-flow gradient, but that such inverse transfer is more efficient for turbulent fluctuations in $3$D than in two dimensions ($2$D).  The transfer is due to linearly stable eigenmodes that are excited nonlinearly. The stable modes, thus, reduce both the nonlinear energy cascade to small scales and the viscous dissipation rate. The vortex-tube stretching is also suppressed. Up-gradient momentum transport by the stable modes counters the instability-driven down-gradient transport, which also is more effective in $3$D than in $2$D ($\mathrm{\approx} 70\% \mathrm{\,\, vs.\,\,}\mathrm{\approx} 50\%$). From unstable modes, these stable modes nonlinearly receive energy via zero-frequency fluctuations that vary only in the direction orthogonal to the plane of $2$D shear flow. The more widely occurring $3$D turbulence is thus inherently different from the commonly studied $2$D turbulence, despite both saturating via stable modes.

\end{abstract}

\maketitle

\section{Introduction}
\label{sec:intro}

The directionality of turbulent energy transfer has long been a topic of interest.\cite{alexakis2018}  Early on, energy transfer to small scales was posited in Navier-Stokes inertial-range turbulence, as part of the scale-invariant property of the nonlinearity.\cite{kolmogorov1941, frisch1995} Later, inviscid dynamical invariants were shown to govern energy transfer directions in two-dimensional ($2$D) turbulence.\cite{kraichnan1967,leith1968}  In $2$D, because the vortex tubes cannot be stretched, two positively signed invariants exist and demand oppositely directed cascades: enstrophy to small and energy to large scales.\cite{kraichnan1967,leith1968}  Nonlinear energy transfer is also affected by global symmetry breaking processes like rotation,\cite{waleffe1993, smith1999, smith2002} stratification,\cite{riley2000} a mean shear flow,\cite{horton2010, mamatsashvili2014, mamatsashvili2016, gogichaishvili2017, gogichaishvili2018, mamatsashvili2020} or an external magnetic field via anisotropic linear physics.\cite{biskamp1995, ng1996, terry2004, du2023}  The important effects observed are spectral condensation at large scales and anisotropic turbulent structures that coincide with the anisotropic linear physics.\cite{rubio2014,guervilly2014}  The mechanism involves the correlation time of nonlinear interactions, which depends on anisotropic frequencies of linear waves.\cite{waleffe1993, smith1999, smith2002, terry2004}  Thus, anisotropy is endowed to, otherwise isotropic, nonlinear energy transfer.  It is by this mechanism that the inverse energy cascades can be explained in $3$D rotating turbulence,\cite{waleffe1993, smith1999, smith2002} and similar arguments apply in the creation of large-scale zonal flows\cite{terry2004} in fusion plasmas where the mean magnetic field breaks the symmetry.  This understanding, however, must account for the physics of the linear instability, when it drives the turbulence.

If the instability operates at large scales,\cite{fraser2017} the tendency of instability to relax its driving gradient implies energy transfer to smaller scales.  This traditional argument anticipates that instability-driven turbulence will feature forward energy transfer.  While true, the linear-instability-supplied fluctuation energy need not necessarily, in its entirety, be available to a nonlinear energy cascade to other scales. A discrepancy can arise because the nonlinearity can excite stable-mode solutions of the operator of the linear stability analysis.\cite{fraser2017, fraser2018, fraser2021, tripathi2022a, tripathi2022b, tripathi2023, terry2006, terry2018, terry2021, makwana2011, whelan2018, makwana2014, hatch2011prl, hatch2011, li2021, li2022} Stable-mode excitation has also been observed in laboratory experiment.\cite{qian2020}   When excited to large amplitudes, the stable modes directly transfer turbulent energy from the instability-scale fluctuations to the driving mean gradient,\cite{tripathi2023} a process that can be interpreted as an inverse transfer of energy. Such direct fluctuation-to-mean transfer is distinct from the nonlinear energy cascade. However, the former can drastically reduce the latter,\cite{tripathi2023, terry2021} a phenomenon that has come to light in turbulence theory in recent years.\cite{terry2004, terry2006}
 
All the effects described above can arise in shear flow-driven hydrodynamic turbulence.  Our chief concerns in this paper relate to energy transfer directionality, spatial dimensionality, and instability saturation mechanism. We investigate the turbulence driven by a forced, unstable shear flow in three dimensions, specifically the Kelvin-Helmholtz (KH) instability. The KH instability is an inviscid instability\cite{chandrashekhar1961} whose eigenspectrum is composed of: (i) discrete eigenfrequencies that are purely imaginary and that appear in complex-conjugate pairs, and (ii) a theoretically infinite number of eigenmodes that form a continuum in real frequency.\cite{tripathi2022b} In this study, the discrete modes are an unstable mode and a damped (conjugate-stable) mode. The former extracts energy from the unstable flow-gradient and the latter reverses it.\cite{fraser2017, tripathi2022b} Being a conjugate-stable mode, its mode structure is conceptually similar to that of the unstable mode.\cite{fraser2021} Because of the similarity in mode structures and complex eigenfrequencies,\cite{terry2006} the conjugate-stable modes can strongly nonlinearly couple to the unstable modes, and thus get excited to large amplitudes.

In $2$D hydrodynamic\cite{fraser2017} and magnetized shear flows,\cite{fraser2021, tripathi2022b, tripathi2022a} the conjugate-stable modes have been shown to be significantly excited in saturation. The modes sequester energy at large scales, reduce the nonlinear energy cascade, and lower the small-scale dissipation rate.\cite{tripathi2022a}  Driving up-gradient momentum transport, the stable modes counter the instability-driven down-gradient transport.\cite{fraser2021,tripathi2022b}

In $3$D fluid turbulence, the strong excitation of conjugate-stable modes remains to be confirmed. Possible reasons for a lack of such a mechanism relate to the added degree of freedom in $3$D, which may result in an inefficient stable-unstable mode coupling, thus allowing the majority of energy to cascade to small scales. However, with the additional dimension, the number of new unstable and conjugate-stable modes (hereafter interchangeably called stable modes) also increases.\cite{chandrashekhar1961}  Since the turbulence is inherently a multi-scale nonlinear phenomenon, with complexity further enhanced by the presence of large-scale energy sources and sinks in the form of unstable modes and stable modes, reliable understanding can only be gained with quantitative assessment. A detailed analysis of the energetics of $3$D turbulence is, thus, crucially needed.

Instability saturation can also be influenced by the anisotropic dispersion relation of the $3$D KH instability, where the anisotropy arises because the perturbations that vary only in the spanwise direction (orthogonal to the direction of the mean flow and the direction of shear) are distinct from the perturbations that vary only in the direction of the mean flow.  When the spanwise perturbations are analyzed in terms of an eigenspectrum, one finds neither the KH instability nor the waves forming a continuum in frequency: all eigenfrequencies of the inviscid linear operator collapse to zero.\cite{chandrashekhar1961} One might, thus, assume that these spanwise fluctuations play no role in the saturation of the KH instability, as the instability is largely caused by the fastest-growing two-dimensional streamwise perturbations.\cite{chandrashekhar1961} However, nonlinear mode couplings can strongly drive the zero-frequency, purely spanwise fluctuations, making the $3$D turbulence different from its $2$D counterpart in essential regards. For example, the spanwise fluctuations, due to their zero frequencies, can catalyze near-resonant\cite{waleffe1993, smith1999, smith2002, terry2004} energy transfer between two waves of similarly high frequencies, as such a triad maximizes the turbulent interaction time\cite{terry2018, terry2021}; in the context of the inviscid-instability-driven dissipative turbulence, this translates to energy transfer between eigenmodes of similar but opposite growth rates.\cite{terry2018, terry2021, li2021, li2022} Hence, this new channel involving spanwise fluctuation can participate in the nonlinear saturation of the instability in $3$D. However, whether the participation becomes dominant is \textit{a priori} unknown.

Even in linear stability analysis, the $2$D and $3$D KH instabilities are different, as shown already in 1933 by Squire, stating\cite{drazin2004} that $2$D perturbations are the least stable. Technically valid only for linear analysis, the theorem is relied upon in insightful nonlinear studies of $2$D turbulence, ignoring the possibility that the $3$D KH-unstable modes may compete nonlinearly with the $2$D KH-unstable modes, and even dominate. A part of the reason is the (computational) cost-effectiveness of the $2$D analysis. Such studies commonly appeal to nonlinearly\cite{smith1999, smith2002} two-dimensionalizing effects such as a strong rotation, although the rotation can introduce new waves.\cite{buzz2018} There have, however, been illuminating studies on $3$D KH-instability-driven turbulence, where fully $3$D fluctuations are connected to the secondary and tertiary instabilities of the fluctuations generated by the primary $2$D KH instability (see, e.g, Refs.~\cite{klaassen1985, klaassen1991, mashayek2012a, mashayek2012b, salehipour2015, smith2021, liu2022}). While such $3$D higher-order instability analyses offer a unique perspective on the development of the $3$D turbulence and characterize the ``zoo" of secondary and tertiary $3$D structures, it can be difficult to extract the importance of the $3$D primary KH mode evolution because such slowly growing KH modes can take a longer time to rise to appreciable levels. Before those are reached, the mean shear flow can decay, causing a premature demise of the $3$D KH modes  The issue is, therefore, not just $3$D, but the time scale that the $3$D primary KH-unstable modes need to be allowed to evolve self-consistently without disadvantaging them by preventing the loss of the instability. A continuously active free energy source for instability can arise in natural and laboratory flows, including those in galaxies, stars, protoplanetary disks, atmospheres, oceans, at the mouths of rivers and estuaries, and nuclear fusion devices, where flow-shear is supplied externally; see, e.g., Refs.~\citep{smith2021, marston2008, cope2020, garaud2016, lucas2017, caulfield2020, garaud2018, fraser2023} In such cases, how the instability saturates nonlinearly is not understood.

Momentum transport by the KH instability is countered by the stable modes:\cite{fraser2017, fraser2018}  the stable modes steepen the mean flow gradient that the KH instability attempts to flatten. Understanding the efficiency of these competing processes allows to construct reduced and truncated models of turbulent transport.\cite{fraser2021, tripathi2022b, terry2021} If the unstable and stable modes primarily contribute to momentum transport in the $3$D KH-instability-driven turbulence, accurate transport models may be achieved. Exploring such possibilities, therefore, benefits geo- and astrophysical communities (see, e.g., Refs.~\cite{masson2018, lecoanet2016}) who wish to understand the turbulence and predict the momentum and particle transport rates due to the $3$D KH instability in nature.

One of the principal findings of this paper is that the $3$D KH instability nonlinearly excites zero-frequency, spanwise fluctuations that near-resonantly transfer energy to the conjugate-stable modes at KH-unstable wavenumbers. These stable modes then return the turbulent energy to the mean flow. Such stable modes inhibit the energy cascade to small scales, viscous dissipation rate, momentum transport across the shear layer, and vortex stretching---a purely $3$D phenomenon.

This paper is organized as follows.  Section~\ref{sec:model} presents the model of shear-flow turbulence and simulation details. In Sec.~\ref{sec:lin-analysis}, a complete eigenmode basis is derived using an inviscid linear operator. Stable-mode excitation is demonstrated in Sec.~\ref{sec:nl-stablemode}. Section~\ref{sec:transport-modeling} shows significant counter-gradient momentum transport by stable modes and presents a reduced transport model. Inverse transfer of energy by the stable modes from the fluctuations to the mean flow is investigated in Sec.~\ref{sec:energetics}. How such stable modes receive energy in the nonlinear phase is explored in Sec.~\ref{sec:near-resonant}. The stable modes are shown in Sec.~\ref{sec:vortex-stretching} to retard $3$D vortex stretching and subdue small-scale dissipation. Section~\ref{sec:discussion} presents broader issues and implications of the results, before conclusions in Sec.~\ref{sec:conclusions}.

\section{Model and simulation setup}
\label{sec:model}

An incompressible hydrodynamic fluid evolves\cite{chandrashekhar1961} according to 
\begin{subequations}
\begin{align}
    \label{eq:dtu}
    &\partial_t \vecu + \vecu \cdot \mathbf{\nabla} \vecu = - \nabla p  + \nu \nabla^2 \vecu + \mathbf{f},\\ 
    \label{eq:divu}
    &\mathbf{\nabla} \cdot \vecu=0,
\end{align}
\end{subequations}
where $\vecu$ is the fluid velocity, $\nu$ the kinematic viscosity, and $\mathbf{f}$ the forcing, or externally supplied acceleration. The pressure (per unit density) $p$ imposes the incompressibility constraint Eq.~\eqref{eq:divu} on the flow.

\subsection{Background flow and forcing}
We consider an initial flow profile $\vecu(x,y,z,t\mathrm{=}0) = U_0\tanh(z/a) \hat{\mathbf{e}}_x$, which is well-known to be unstable.\cite{chandrashekhar1961} The flow amplitude $U_0$ and half-width $a$ of the flow-shear are the initial characteristic speed and length scale that are used to non-dimensionalize all the variables henceforth.  Thus, time is measured in units of $a/U_0$ and energy (per unit mass) in terms of $U_0^2$. The initial mean flow thus becomes $\vecu(x,y,z,t\mathrm{=}0) = \tanh(z) \hat{\mathbf{e}}_x$. The viscosity is non-dimensionally measured via the Reynolds number $Re\mathrm{=}U_0 a/\nu$.

In natural systems, the unstable mean shear flow can evolve in at least two different qualitative ways: first, when the mean shear flow is allowed to freely evolve following the response from the shear flow-driven turbulence, the unstable shear flow relaxes to a stable configuration. Second, when the mean shear flow is allowed to evolve in the presence of the external drive that brings the shear flow in existence in the first place, the resulting turbulence can be of quasi-stationary nature. Admittedly, the form of the external drives can vary from system to system; avoiding such a system-specific requirement, we use a general form of external forcing that replenishes the initial, unstable equilibrium. Such type of restoring force have previously been used to model external forces in natural environments,\cite{marston2008, smith2021, cope2020, pueschel2011, pueschel2014} e.g., jets in stars, driven flows in geo-and astrophysics, which can be related to various forms of external drives,\cite{smith2021, marston2008, cope2020, garaud2016, lucas2017, caulfield2020} for example, fingering convection that ceaselessly seeds secondary instability with shear flow,\cite{garaud2018, fraser2023} and, in general, the gravitational force that generates shear layers in accretion disks and stellar interiors. We thus externally drive the mean flow using $\mathbf{f} = f(k_x\mathrm{=}0, k_y\mathrm{=}0, z, t) \hat{\mathbf{e}}_x$, where $k_x$ and $k_y$ are the wavenumbers along the $x$- and $y$-axes, respectively, with
\begin{equation}
    f = \frac{1}{t_\mathrm{forcing}} \left[ U_\mathrm{ref}(z) - \left\langle u_x(x,y,z,t)\right\rangle_{x,y}\right] + F_0,
\end{equation}
where $U_\mathrm{ref}(z)\mathrm{=}\tanh(z)$ is the initial mean flow; $\left\langle u_x(x,y,z,t)\right\rangle_{x,y}$ is the $(x,y)$-averaged instantaneous flow, directed along the $x$-axis; and $1/t_\mathrm{forcing}$, with units of $U_0/a$, is the rate at which the mean flow is forced toward the initial equilibrium profile.\cite{marston2008, smith2021, tripathi2022b}  A time-independent forcing $F_0$ helps to realize a true equilibrium of the initial mean flow: $Re^{-1}\nabla^2 U_\mathrm{ref}(z) + F_0 = 0$, allowing a formal linear analysis.

On this equilibrium system, small-amplitude initial noise is seeded for the present simulation studies.

\subsection{Initial perturbations, boundary conditions, and simulation parameters}

To ensure incompressiblity $\mathbf{\nabla} \cdot \vecu\mathrm{=}0$ of the flow in the initial random perturbations, careful steps are taken by first writing $\vecu \mathrm{=}\nabla \times \mathbf{C}$, where $\mathbf{C} \mathrm{=} \sum_{\mu\mathrm{=}\{x,y,z\}} C_\mu \hat{\mathbf{e}}_\mu$ is a $3$-dimensional vector field, with each component given as
\begin{equation}\label{eq:pertform2}
C_\mu(x,y,z) = \sum_{\substack{0 < |k_x| < k_x^\mathrm{max}\\ 0 < |k_y| < k_y^\mathrm{max}}} \mathrm{e}^{i k_x x} \mathrm{e}^{i k_y y}  \hat{C}_\mu(k_x,k_y,z).
\end{equation}
Initial perturbations are excited up to $k_x = k_x^\mathrm{max}$ and $k_y = k_y^\mathrm{max}$. The Fourier components of such perturbations are
\begin{equation}\label{eq:pertform}
\hat{C}_\mu(k_x,k_y,z) =  \hat{\mathbf{e}}_\mu A\, \mathrm{e}^{-\frac{z^2}{\sigma^2}}  \Bigg(  |k_x|^\alpha \mathrm{e}^{i \theta_\mu(k_x)} \delta_{k_y,0} 
+ |k_y|^\alpha \mathrm{e}^{i \theta_\mu(k_y)} \delta_{k_x,0}
+  |k_x k_y|^\beta  \mathrm{e}^{i \theta_\mu(k_x,k_y)}\Bigg),
\end{equation}
where $A$ is an overall amplitude of the perturbation (measured non-dimensionally with respect to the initial mean-flow amplitude $U_0$); $\sigma$ is the width (measured non-dimensionally with respect to the half-width of the flow-shear $a$) along the $z$-axis of the Gaussian envelope that scales the initial perturbations; $\alpha$ and $\beta$ are the spectral slopes of the perturbations in the wavenumber space $k_x$ and $k_y$, respectively; each component of $\mathbf{C}$ is randomized in phase at every wavenumber via random phase functions $\theta_\mu(k_x), \theta_\mu(k_y),$ and $\theta_\mu(k_x, k_y)$. On the right-hand side of Eq.~\eqref{eq:pertform}, the first term inside the brackets excites the $k_y\mathrm{=}0$ Fourier modes, which are the purely $2$D perturbations; the second term corresponds to the fluctuations that vary only on the $(y,z)$-plane, orthogonal to the streamwise direction; and the third term excites perturbations that vary in all directions.

Motivated by earlier studies of $2$D shear-flow turbulence,\cite{fraser2021,tripathi2022a,tripathi2022b,tripathi2023} we use a small-amplitude perturbation, with $A\mathrm{=}10^{-3}$, to allow an unambigious and decades-long linear evolution of the instability, before it gets nonlinearly modified. The $2$D studies employed only the first term inside the brackets on the right-hand side of Eq.~\eqref{eq:pertform}. The initially excited perturbations, up to $k_x^\mathrm{max}=128$ and $k_y^\mathrm{max}=128$, have a near-flat spectrum of energy over the ($k_x, k_y$)-plane, which we achieve by choosing $\alpha\mathrm{=}-1$ and $\beta\mathrm{=}-1/2$ in Eq.~\eqref{eq:pertform}.  The perturbations peak at $z\mathrm{=}0$ and have a Gaussian width of $\sigma\mathrm{=}2$.

To make the initial perturbations fully free of $\vecu$-divergences, the pressure $p$ is obtained using $p \mathrm{=} \nabla^{-2}\nabla\cdot\left[-(\vecu\cdot\nabla)\vecu \right]$ where $\nabla^{-2}$ stands for inversion of the Laplacian operator.

Boundary conditions along the $x$- and $y$-axes are periodic, whereas, along the $z$-axis, we impose no-slip, co-moving (with the initial flow) walls at the top and bottom boundaries $z\mathrm{=}\pm L_z/2$.

While we have studied the current system with varying forms of initial condition, box size, spectral resolution, and the Reynolds number, we present in this paper results using a box size of $L_x\mathrm{=}L_y\mathrm{=}L_z\mathrm{=}10\pi$, resolving a large number of linearly KH-unstable wavenumbers, and using a spectral resolution of $512^3$: Fourier modes on the $(x,y)$-plane and Chebyshev polynomials along the $z$-axis.  To avoid aliasing errors, we dealias the quadratic nonlinearities of the governing equation via the $3/2$-rule in all directions. Motivated by earlier studies,\cite{tripathi2022a, tripathi2022b, tripathi2023} $t_\mathrm{forcing}\mathrm{=}0.5$ is chosen throughout this study (the instability-growth time is $\sim 6$ and the box-crossing time of the mean flow is $10\pi $). Such a choice adequately replenishes the initial mean flow, thereby preventing the loss of the KH instability. We use the pseudospectral solver Dedalus,\cite{burns2020} with third-order, four-stage implicit-explicit Runge-Kutta time-integrator to advance the system. At the cost of $\mathrm{\sim}10$ million CPU-hours, long time-integrated $3$D simulations are studied for $Re=300, 1500,$ and $5000$, in addition to an ensemble of $10$ different three-dimensional numerical simulation continuations (in Sec.~\ref{sec:vortex-stretching}).

\section{Complete eigenmode basis}
\label{sec:lin-analysis}

Before analyzing the nonlinear state of the shear-flow instability, we find the inviscid eigenmodes by linearizing Eqs.~\eqref{eq:dtu} and \eqref{eq:divu}, dropping the viscous term, and Fourier-transforming the equations,
\begin{subequations}
\begin{align}
    \label{eq:dtulin}
    &\partial_t \hat{\vecu} + \vecu_0 \cdot \mathbf{\nabla} \hat{\vecu} + \hat{\vecu} \cdot \mathbf{\nabla} \vecu_0 = - \nabla \hat{p},\\ 
    \label{eq:divulin}
    &\mathbf{\nabla} \cdot \hat{\vecu}=0,
\end{align}
\end{subequations}
where $\vecu$ is decomposed into its mean ($k_x\mathrm{=}k_y\mathrm{=}0$) mode $\vecu_0$ and fluctuation spectrum $\hat{\vecu}$ at $(k_x,k_y)$; likewise for the pressure $\hat{p}$; the derivative operator $\nabla$ can be expressed as $i k_x \hat{\mathbf{e}}_x + i k_y \hat{\mathbf{e}}_y + \hat{\mathbf{e}}_z \partial_z$.  

\begin{figure*}
    \centering
    \includegraphics[width=0.9\textwidth]{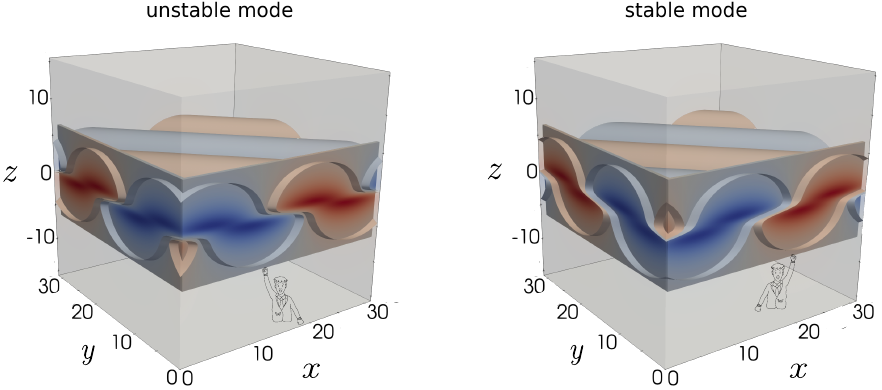}
    \caption{For an unstable mean flow $\vecu_0 \mathrm{=} \tanh(z) \hat{\mathbf{e}}_x$, physical structures of perturbed $\uz$ of $3$D unstable and conjugate-stable modes are shown; the colors represent the magnitude of the $\uz$ (in arbitrary units). The isocontours and the surface rendering highlight a symmetry that is illustrated with a cartoon.}
    \label{fig:f1}
\end{figure*}

With $\vecu_0 \mathrm{=} U_\mathrm{ref}(z) \hat{\mathbf{e}}_x$, a complete eigenspectrum can be found at any given wavenumber.  When $k_y\mathrm{=}0$, one obtains an eigenspectrum corresponding to $2$D perturbations:\cite{tripathi2022b} (i) unstable and conjugate-stable eigenmodes at $|k_x|\lesssim 1$, frequencies of which are complex-conjugates, and (ii) a large (theoretically infinite) number of eigenmodes that form a continuum in real frequency and that exist at all $k_x \mathrm{\neq} 0$.  This set of eigenmodes also exists for $3$D perturbations.  To obtain the eigenmodes, we discretize Eqs.~\eqref{eq:dtulin} and \eqref{eq:divulin} using $1024$ Chebyshev polynomials for each variable ($\ux, \uy, \uz, \hat{p}$) along the $z$-axis, and form a $4096\times 4096$-sized matrix eigenvalue problem at each wavenumber, and then diagonalize the matrix to find the complete eigenspectrum. The $3$D unstable and stable eigenmodes are visually very similar as can be seen, for example, when examining their $z$-components of velocity at the first Fourier mode numbers along the $x$- and $y$-axes in Fig.~\ref{fig:f1}.  The similarity appears because, in wavenumber space, they exhibit a special complex-conjugate symmetry: $(\ux,\uy,\uz,p)_\mathrm{stable\ mode} = (\ux^\ast,\uy^\ast,-\uz^\ast,p^\ast)_\mathrm{unstable\ mode}$, where ${}^\ast$ denotes the complex-conjugation operation.

\begin{figure*}
    \centering
    \includegraphics[width=0.49\textwidth]{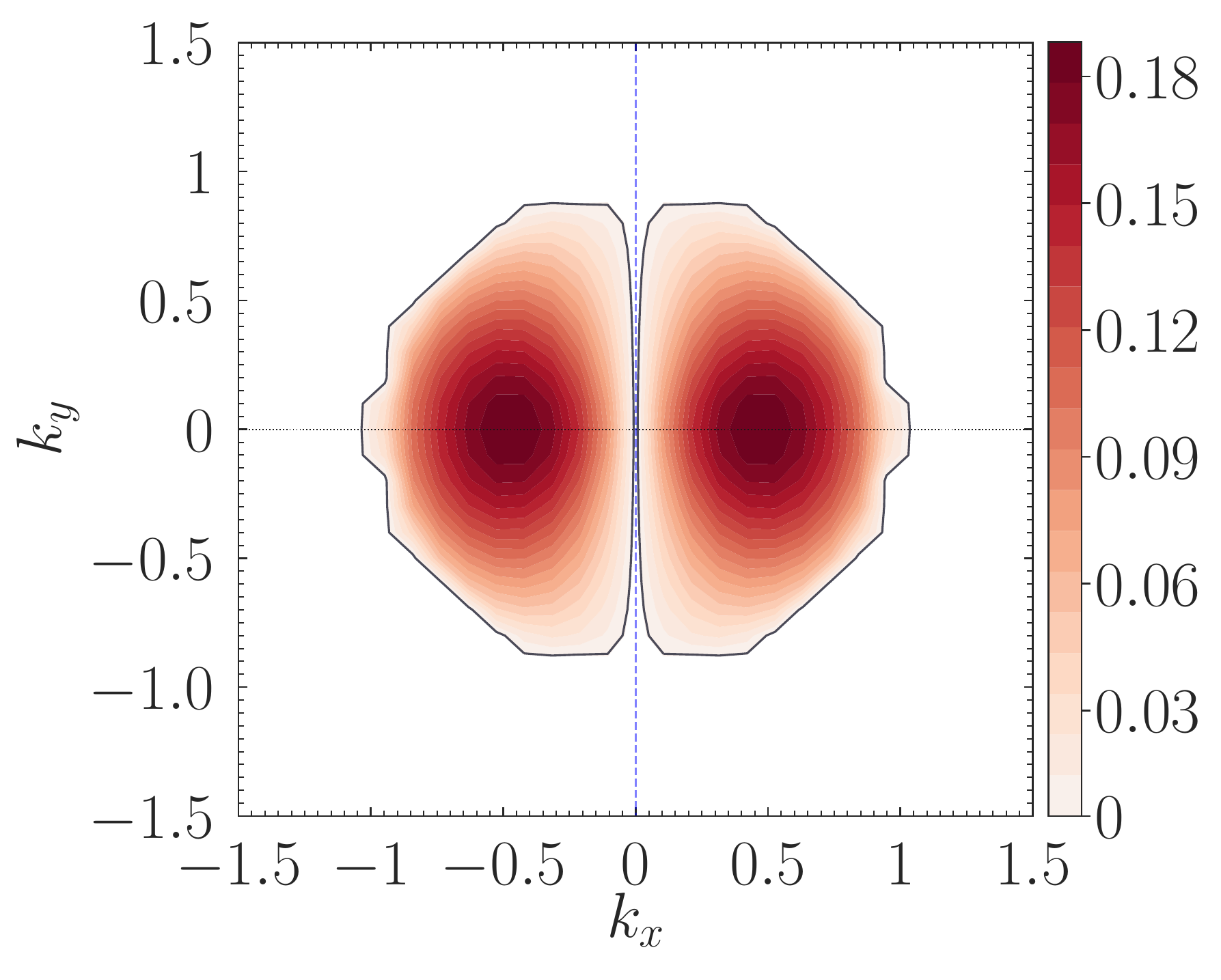}
    \caption{Growth-rate spectrum of $3$D KH instability, showing an anisotropic dispersion relation. The $2$D modes, $k_y\mathrm{=}0$, have the largest growth rates. All wavenumbers with $k_x\mathrm{=}0$, shown with a vertical blue dashed line, are linearly stable.  The wavenumber $(0,0)$ corresponds to the mean flow. As the Kelvin-Helmholtz instability is a large-scale instability, the wavenumbers shown in white, whose boundary is approximated by the black bounding curve, are linearly stable.}
    \label{fig:f2}
\end{figure*}

A growth rate spectrum of the unstable mode on the ($k_x, k_y$)-plane is shown in Fig.~\ref{fig:f2}. At every wavenumber for which the unstable mode of the inviscid linear operator exists, the conjugate-stable mode must necessarily, as well exist---meaning that Fig.~\ref{fig:f2} also serves as a damping rate spectrum of the conjugate-stable mode.  For the special case of $k_x\mathrm{=}0$, the entire eigenspectrum becomes infinitely degenerate, i.e., all eigenvalues collapse to zero on the complex-frequency plane.  In that case, the dropped viscous term becomes singularly important, and is the sole active term that damps the fluctuation spectrum.

\section{Nonlinear excitation of stable modes}
\label{sec:nl-stablemode}

Turning now to the nonlinear saturation of the instability, at any wavenumber $\mathbf{k} \mathrm{=}(k_x, k_y)$, where discrete linear eigenfrequencies in this system exist, an arbitrary Fourier-transformed state vector $\hat{\chi} \mathrm{=} (\ux, \uy,\uz,\hat{p})$ can be decomposed in the basis spanned by the complete set of eigenmodes in the manner of $\hat{\chi} \mathrm{=} \sum_j \beta_j \hat{\chi}_j$, where $\beta_j$ is the complex-valued amplitude of the $j$th eigenmode $\hat{\chi}_j$. Since the eigenmodes at hand are non-orthogonal, the mode-amplitudes $\beta_j$ in the turbulent state cannot be determined simply by taking a Hermitian inner product between the eigenmodes.  However, this is merely a technical challenge; the $\beta_j$ can be uniquely determined by employing the biorthogonal basis, found by solving an adjoint of Eqs.~\eqref{eq:dtulin} and \eqref{eq:divulin}, which produces the same eigenspectrum but a different set of eigenmodes, usually known as left eigenmodes or adjoint solutions.  To gain machine-precision accuracy,\cite{tripathi2022b, paul2019} we use adjoint solutions of discretized equations rather than that of the analytical equations.

\begin{figure*}
    \centering
    \includegraphics[width=0.99\textwidth]{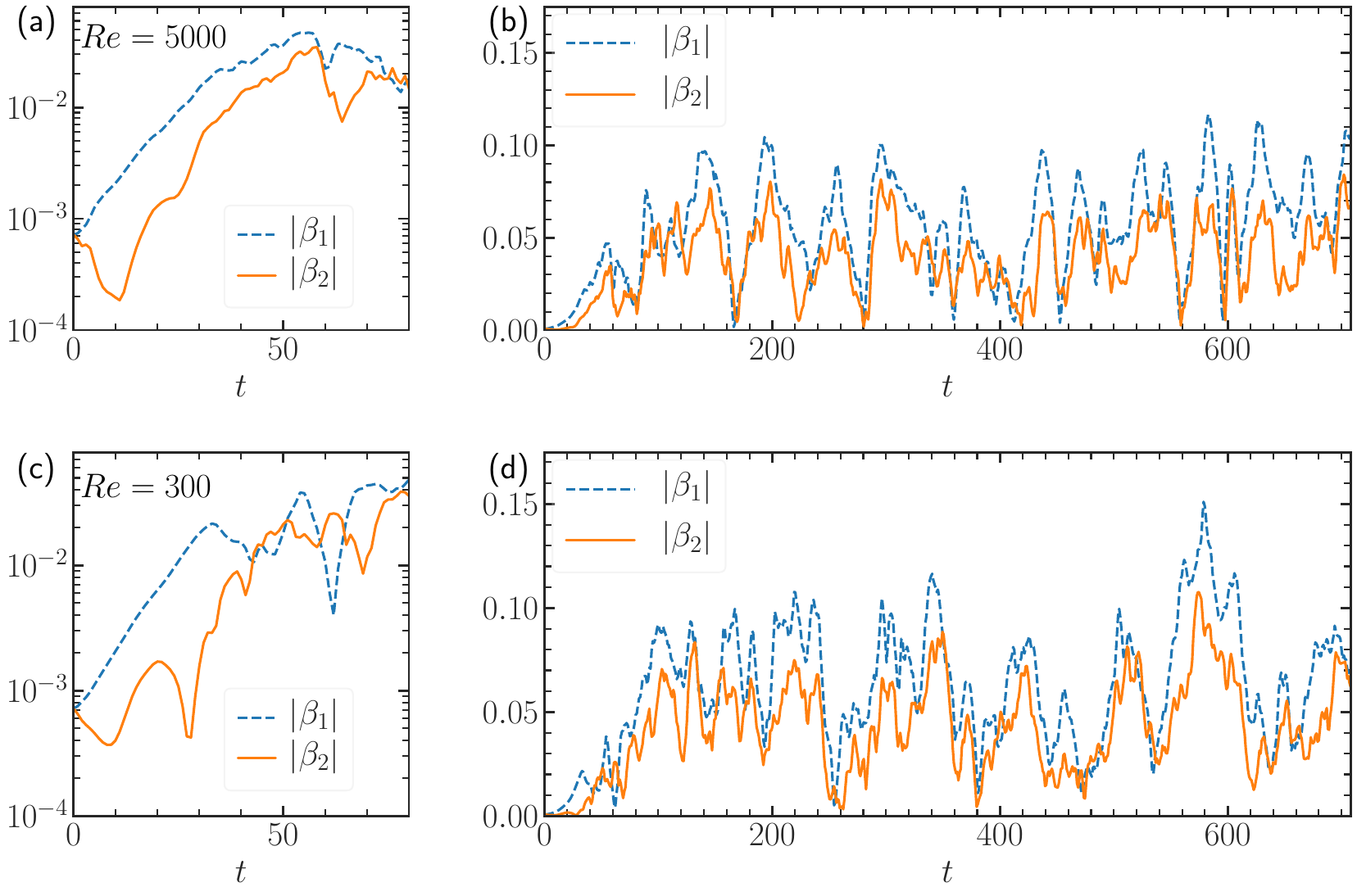}
    \caption{Amplitude evolution of unstable ($|\beta_1|$) and conjugate-stable ($|\beta_2|$) modes at $k_x\mathrm{=}k_y\mathrm{=}0.2$. In (a) and (b), $Re\mathrm{=}5000$ is used in the initial value simulation; and in (c) and (d), $Re\mathrm{=}300$ is used. In each case, the left panel shows the early evolution using a logarithmic vertical scale, whereas the right panel displays the full time evolution using linear scales on both axes. Stable modes are nonlinearly excited much before the saturation of the unstable mode.}
    \label{fig:f3}
\end{figure*}

As the unstable modes grow exponentially in the linear phase, they attain large amplitudes. Then the unstable modes, via nonlinear process, excite eigenmodes that are linearly stable (i.e., decaying in the absence of nonlinearity), as shown with the orange curves in Figs.~\ref{fig:f3}(a) and \ref{fig:f3}(c). The nonlinear interaction quickly involves multiple eigenmodes and multiple wavenumbers, and, in the fully nonlinear phase, in Figs.~\ref{fig:f3}(b) and ~\ref{fig:f3}(d), the conjugate-stable mode attains an amplitude nearly identical to that of the unstable mode. These figures show that the stable modes are excited even in $3$D.  Their excitation is universal, as long as the viscosity is not enormously large to lead to a laminar flow. In support of this, we find, for $Re=1500$ and other lower $Re$ cases (not shown), large excitation of stable modes.

\section{Efficient $3$D counter-gradient momentum transport}
\label{sec:transport-modeling}
We inquire if momentum is transported differently across the mean shear layer due to the stable-mode-rich turbulence. Prior studies have shown counter-gradient momentum transport due to stable modes, but only in $2$D systems,\cite{fraser2021, tripathi2022b, terry2009} leaving the question open whether such a result is generalizable to $3$D systems. To address this issue, we measure the Reynolds stress by decomposing the turbulent fluctuations at each KH-unstable wavenumber into eigenmodes and by evaluating stress contribution from each eigenmode. Since the momentum is transported maximally at the middle of the shear layer $z\mathrm{=}0$, because of the strongest shear there, we compare the transport rates at $z\mathrm{=}0$ from individual eigenmodes, along with the total transport from mode-undecomposed fluctuations at $z\mathrm{=}0$.

\begin{figure*}
    \centering
    \includegraphics[width=0.99\textwidth]{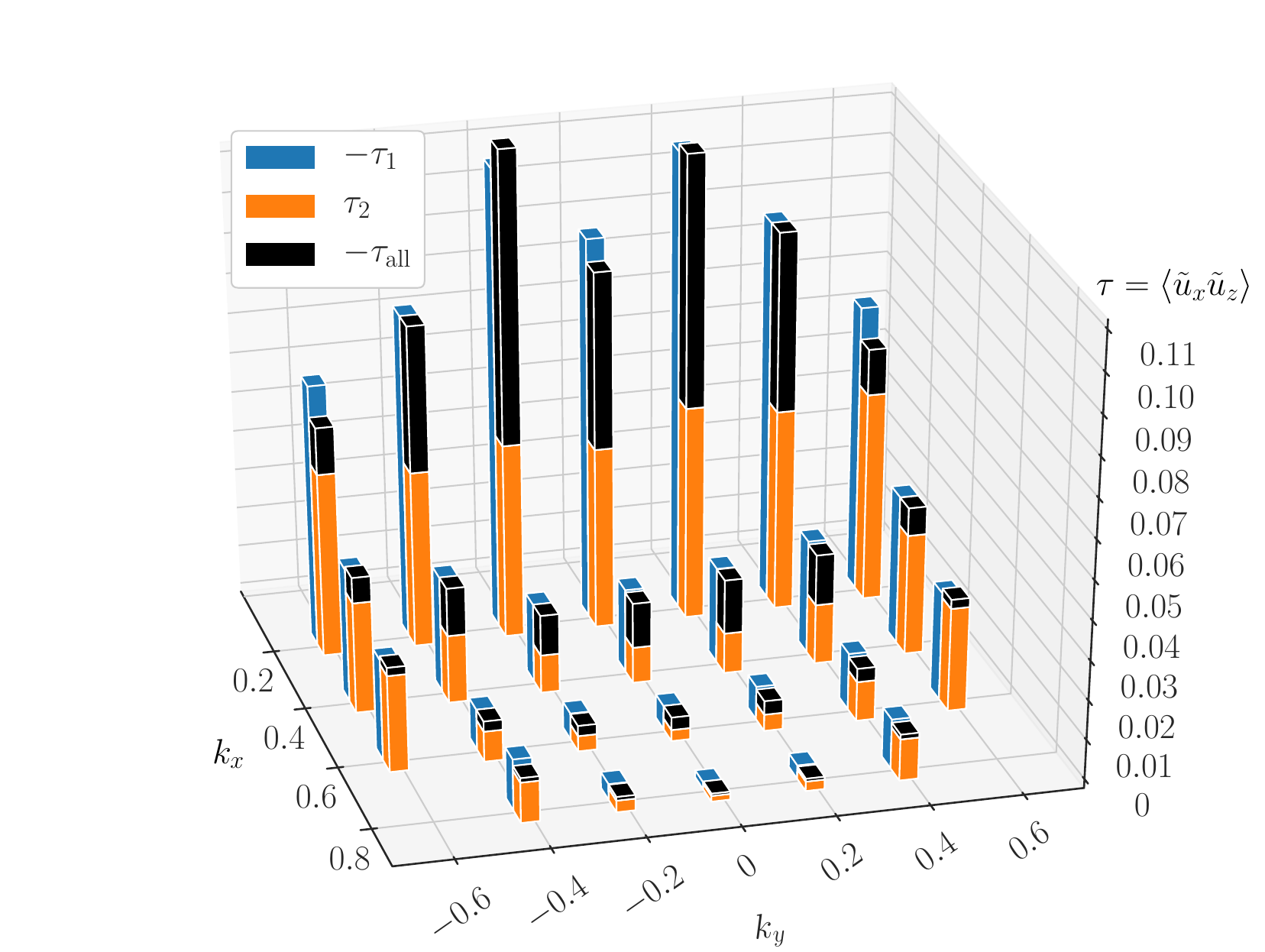}
    \caption{Time-averaged spectrum of Reynolds stress due to: the unstable mode $\tau_1$, stable mode $\tau_2$, and mode-undecomposed full fluctuations $\tau_\mathrm{all}$. Where the peaks of the blue and black bars are at similar heights, $\tau_1+\tau_2$ is a good approximation to $\tau_\mathrm{all}$. The difference between $\tau_1+\tau_2$ and $\tau_\mathrm{all}$ is small, and arises from the contribution of the continuum modes to the total stress $\tau_\mathrm{all}$. All stresses are averaged over $t\mathrm{=}200\textrm{--}708$ for $Re\mathrm{=}5000$.}
    \label{fig:f4n}
\end{figure*}

The time-averaged spectrum of the Reynolds stress is shown in Fig.~\ref{fig:f4n}: in blue, the stress from an unstable mode; in orange, the stress from a stable mode; and in black, the total stress without performing the eigenmode decomposition. The total stress $\tau_\mathrm{all}(z\mathrm{=}0)$ is large (the black bars are taller) for low wavenumbers. Mode-decomposition shows that transport from $3$D unstable modes is large, and correspondingly large is that from the $3$D stable modes.  This immediately challenges a prediction of transport based on unstable modes only.\cite{fuller2019, pessah2006, goodman1994, garaud2018, barker2019} A correct transport level, however, can be found by simply adding the contribution of stable modes.

\begin{figure*}
    \centering
    \includegraphics[width=0.55\textwidth]{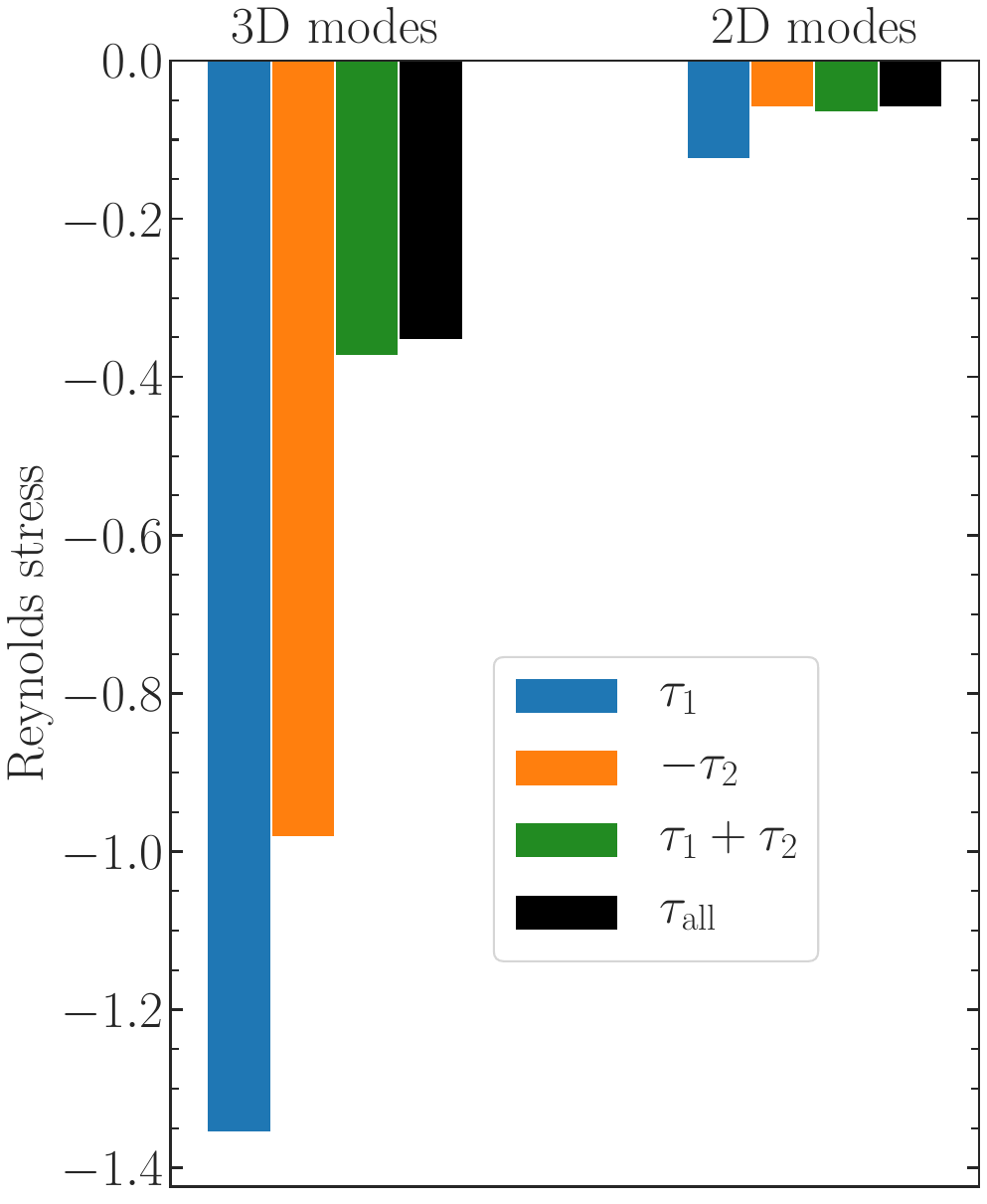}
    \caption{Comparison of time-averaged Reynolds stresses at $z\mathrm{=}0$ for all KH-unstable $3$D vs. $2$D modes. Further decomposition shows contributions from the unstable mode $\tau_1$; stable mode $\tau_2$; the two added together, $\tau_1+\tau_2$; and the mode-undecomposed full fluctuations $\tau_\mathrm{all}$. The difference between $\tau_\mathrm{all}$ and $\tau_1+\tau_2$ arises from the stress due to the continuum modes.  Stresses are averaged over $t\mathrm{=}200\textrm{--}708$ for $Re\mathrm{=}5000$.}
    \label{fig:f5n}
\end{figure*}

To separate the $2$D and $3$D modes, we apply a wavenumber-summation to Fig.~\ref{fig:f4n} and obtain a $2\mathrm{D-}3\mathrm{D}$ comparison in Fig.~\ref{fig:f5n}. It is clear that the total stress, shown in black, is higher for the $3$D modes than for the $2$D modes.  This is easily explained with the aid of Fig.~\ref{fig:f4n} where many $3$D modes have comparable stress contribution to that of the $2$D modes, in addition to the higher density of states of the $3$D modes, i.e., number of wavenumbers $k_y$ per $k_x$. 

\section{Inverse transfer of energy in $3$D}\label{sec:energetics}

We now analyze the energy transfer rate $Q_1(\mathbf{k})$ from the mean flow to the fluctuations at $\mathbf{k}$ by the unstable mode, and the rate of inverse energy transfer $Q_2(\mathbf{k})$ from the fluctuation $\mathbf{k}$ to the mean flow by the stable mode.  Both of these are linear processes, as one of the involved wavenumbers is the mean flow---the reservoir of the free energy for the instability and thus turbulence. Two additional linear processes, however, exist: first, because the mean flow also evolves, the instantaneous deviation of the mean from the initial profile can induce a (typically small) linear coupling between the inviscid eigenmodes of the initial profile, thus reducing $Q_j$ by an amount $R_j$. Second, the viscous term introduces energy dissipation $D_j$. However, since the unstable and stable eigenmodes that are analyzed here exist only at large scales ($|k| \lesssim 1$), the dissipative effect at such scales is small, compared to $Q_j$.  This remains true as long as the viscosity is not enormously large to stabilize completely the inviscid Kelvin-Helmholtz (KH) instability; such a case, however, is irrelevant to the scenario we are interested in, where the KH instability drives the turbulence.

To quantify the energy transfer rates described in the previous paragraph, we Fourier-transform Eqs.~\eqref{eq:dtu} and \eqref{eq:divu}, and project the equation onto an eigenmode of interest, say the $j$th eigenmode, by multiplying the Fourier-transformed equation with the $j$th modified left eigenmode $\hat{Z}_j$; see Appendix~A of Ref.~\cite{tripathi2022b} for a general overview of modified left eigenmodes.  To illustrate this here, Eqs.~\eqref{eq:dtu} and \eqref{eq:divu} can be structurally written, at $\mathbf{k}\mathrm{\neq}(0,0)$, as
\begin{equation} \label{eq:structuraleqn}
    \partial_t M \hat{\chi} = L \hat{\chi} + \sum_{\mathbf{k}', \mathbf{k}'': \mathbf{k}'+\mathbf{k}''=\mathbf{k}}  N(\hat{\chi}', \hat{\chi}''),
\end{equation}
where the matrix $M$ of size $4\times 4$ has the form $[[\mathbb{I}_{3\times 3}, 0],[0,0]]$, with $\mathbb{I}_{3\times 3}$ as the identity matrix of size $3\times 3$; furthermore, $L$ is a linear operator and $N$ a nonlinear operator, with $\hat{\chi}'$ and $\hat{\chi}''$ representing state vectors at $\mathbf{k}'$ and $\mathbf{k}''$, respectively. 

Projecting Eq.~\eqref{eq:structuraleqn} on the modified left eigenmode $\hat{Z}_j$, we obtain the mode-amplitude evolution equation
\begin{equation} \label{eq:dtbetaj}
    \partial_t \beta_j(\mathbf{k}) = \gamma_j(\mathbf{k}) \beta_j(\mathbf{k}) + \sum_{l}B_{jl}(\mathbf{k}) \beta_l(\mathbf{k}) + \sum_{\substack{\mathbf{k}', \mathbf{k}'':\\ \mathbf{k}'+\mathbf{k}''=\mathbf{k}}}\sum_{m,n} C_{jmn}(\mathbf{k}, \mathbf{k}') \beta_m' \beta_n'',
\end{equation}
where we have used the biorthogonality relation between the modified left eigenmode $\hat{Z}_j$ and the right eigenmode $\hat{\chi}_l$ such that $\langle \hat{Z}_j, \hat{\chi}_l\rangle \propto \delta_{j,l}$. In Eq.~\eqref{eq:dtbetaj},  $\gamma_j(\mathbf{k})$ is the complex-valued growth rate of the $j$th eigenmode at a wavenumber $\mathbf{k}$; the coefficient $B_{jl}(\mathbf{k})$ measures the linear coupling (between the inviscid eigenmodes at $\mathbf{k}$) induced by the viscous term and by the instantaneous deviation of the mean flow from the initial profile; the coefficient $C_{jmn}(\mathbf{k}, \mathbf{k}')$ measures the nonlinear coupling between the eigenmode $m$ at $\mathbf{k}'$ and the eigenmode  $n$ at $\mathbf{k}''$, driving the eigenmode $j$ at $\mathbf{k}$; and $\beta_m'$ and $\beta_n''$ represent the amplitudes of modes $m$ at $\mathbf{k}'$ and $n$  at $\mathbf{k}''$, respectively. Now, we multiply Eq.~\eqref{eq:dtbetaj} with $\beta_j^\ast$, and add the complex-conjugate of the resulting equation to obtain the evolution equation for the eigenmode energy, which reads
\begin{equation} \label{eq:dtbetajenergy}
    \partial_t |\beta_j(\mathbf{k})|^2 = Q_j + \left(R_j + D_j \right) + T_j,
\end{equation}
where
\begin{subequations}
\begin{align}
    Q_j &= 2\real\left[\gamma_j(\mathbf{k})\right] |\beta_j(\mathbf{k})|^2,\\
    R_j+D_j &= 2\real\Big[\sum_{l}B_{jl} \beta_l \beta_j^\ast\Big],\\  \label{eq:Tj}
    T_j &=  \sum_{\substack{\mathbf{k}', \mathbf{k}'':\\ \mathbf{k}'+\mathbf{k}''=\mathbf{k}}}\sum_{m,n}  T_{jmn}(\mathbf{k}, \mathbf{k}'),\\
    T_{jmn}(\mathbf{k}, \mathbf{k}') &= 2\real [C_{jmn}(\mathbf{k}, \mathbf{k}') \beta_m' \beta_n'' \beta_j^\ast].
\end{align}
\end{subequations}
In Eq.~\eqref{eq:dtbetajenergy}, $Q_j$ measures the energy transfer rate by the $j\mathrm{th}$ eigenmode to the fluctuation scale $\mathbf{k}$ from the initial mean flow; $R_j$ is the energy transfer rate by the $j\mathrm{th}$ eigenmode to the fluctuation scale $\mathbf{k}$, the energy source here being the instantaneous deviations of the mean flow from the initial mean flow; $D_j$ is the viscous dissipation rate by the $j\mathrm{th}$ eigenmode at the fluctuation scale $\mathbf{k}$; $T_j$ is the nonlinear drive to the $j\mathrm{th}$ eigenmode. Since the eigenmodes used for the basis expansion are the inviscid eigenmodes of the initial mean flow, the terms $R_j$ and $D_j$ introduce a linear coupling $B_{jl}$ between those eigenmodes, the effect of which we measure next.

\begin{figure*}
    \centering
    \includegraphics[width=0.9\textwidth]{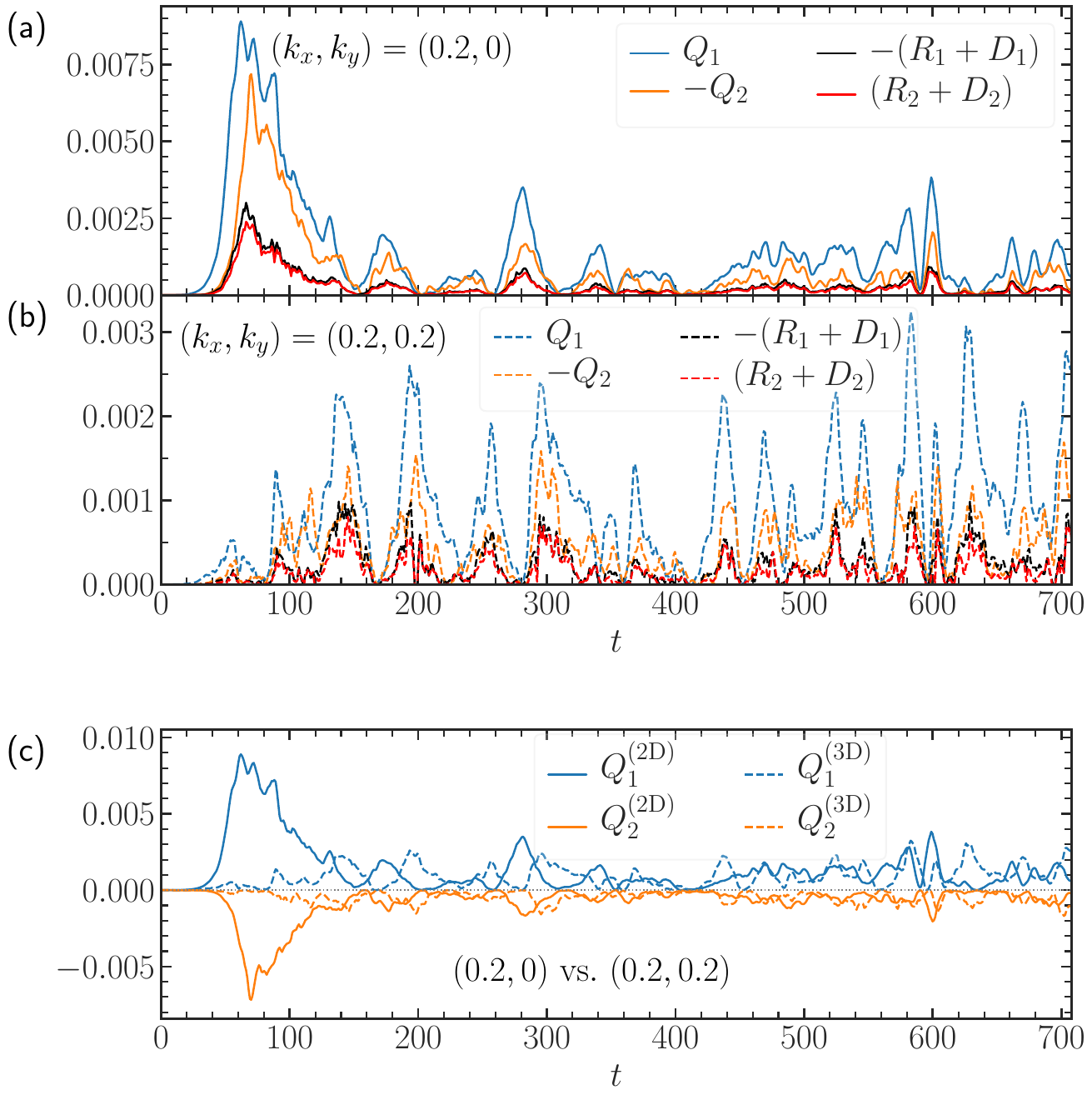}
    \caption{Time traces of energy transfer to an unstable, $j\mathrm{=}1$, and a stable, $j\mathrm{=}2$, mode at wavenumbers with (a) $2$D and (b) $3$D perturbations, in a simulation with $Re\mathrm{=}5000$. The quantity $Q_j$ denotes the energy transfer rate from the initial mean flow to the $j\mathrm{th}$ eigenmode at fluctuation scale $(k_x, k_y)$; $R_j\mathrm{+}D_j$ collectively represents the effect of time-deviations of the instantaneous mean flow from the initial profile and the effect of viscous dissipation. In (c), $2$D and $3$D perturbations, decomposed by unstable and stable modes, are compared in terms of the energy transfer rate $Q_j$; the stable modes reverse the fluctuation energy to the mean flow at all times, once the nonlinearity excites them.}
    \label{fig:f6n}
\end{figure*}

For both the $2$D and $3$D perturbations that the system generates, the energy transfer rates $Q_j$ and $R_j+D_j$ are shown in Figs.~\ref{fig:f6n}(a) and \ref{fig:f6n}(b) for the energetically dominant wavenumbers. The energy transfer reduction term $R_j + D_j$ is of opposite sign compared to $Q_j$. This physically means that unstable modes are slightly less efficient in extracting energy from the instantaneous mean flow when compared with their efficiency of energy extraction from the initial mean flow; the same applies to the energy return rate by the stable modes. This particular finding is similar to the results reported for $2$D MHD shear-flow turbulence.\cite{tripathi2023}

At early times $t\lesssim 140$, the $2$D mode dominates over the $3$D mode; Fig.~\ref{fig:f6n}(c). However, in the turbulent phase, the slowly growing $3$D wavenumber competes with the $2$D wavenumber. In all cases, the stable modes are significantly excited, nearly to the levels of the corresponding unstable modes. 

We note that, because of the added degree of freedom (the $y$-coordinate) in the $3$D system, there exist more wavenumbers that are $3$D KH-unstable than that are $2$D KH-unstable; for example, there is only one wavenumber that is $2$D KH-unstable at $k_x\mathrm{=}0.2$, i.e., $(0.2, 0)$, whereas there are multiple wavenumbers that are $3$D KH-unstable at $k_x\mathrm{=}0.2$, namely, $(0.2, \pm 0.2)$, $(0.2, \pm 0.4)$, etc. We avoid double-counting the Hermitian conjugates ($-k_x, -k_y$) of these wavenumbers ($k_x, k_y$), which are already included in the definition of $Q_j(\mathbf{k})$. 

\begin{figure*}
    \centering
    \includegraphics[width=0.99\textwidth]{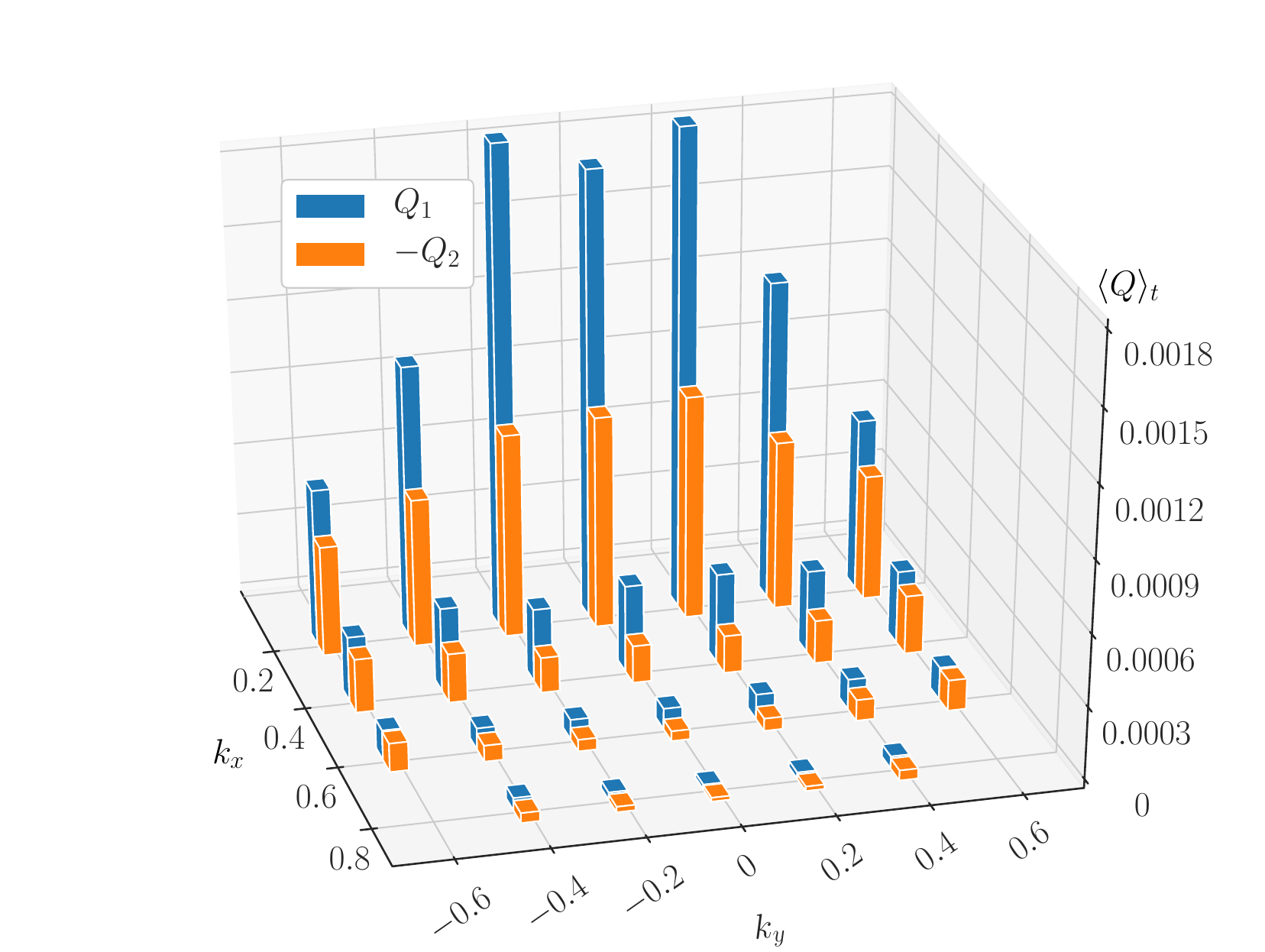}
    \caption{Time-averaged rate of energy transfer from initial mean to fluctuation $Q_1$ by unstable modes, and inverse transfer $Q_2$ by stable modes, in a simulation with $Re\mathrm{=}5000$; the time-averaging interval is $t\mathrm{=}200\textrm{--}708$. Because of the Hermitian conjugacy in the perturbations over the wavenumbers, only a non-redundant half of the $(k_x, k_y)$-plane is shown. The linearly fastest growing mode exists at $\mathbf{k}\mathrm{=}(0.4, 0)$.}
    \label{fig:f7n}
\end{figure*}

Time-averaged energy-transfer rates are compared in Fig.~\ref{fig:f7n}, where the rates are decomposed by unstable and stable modes. Despite growing slowly, the $3$D wavenumbers contribute significantly in the fully nonlinear phase in extracting energy from the mean flow.  However, the stable modes at the same wavenumbers, whether $2$D or $3$D, always efficiently deplete fluctuation energy and transfer energy in the reverse direction: from the fluctuation to the mean flow.  

\begin{figure*}
    \centering
    \includegraphics[width=0.5\textwidth]{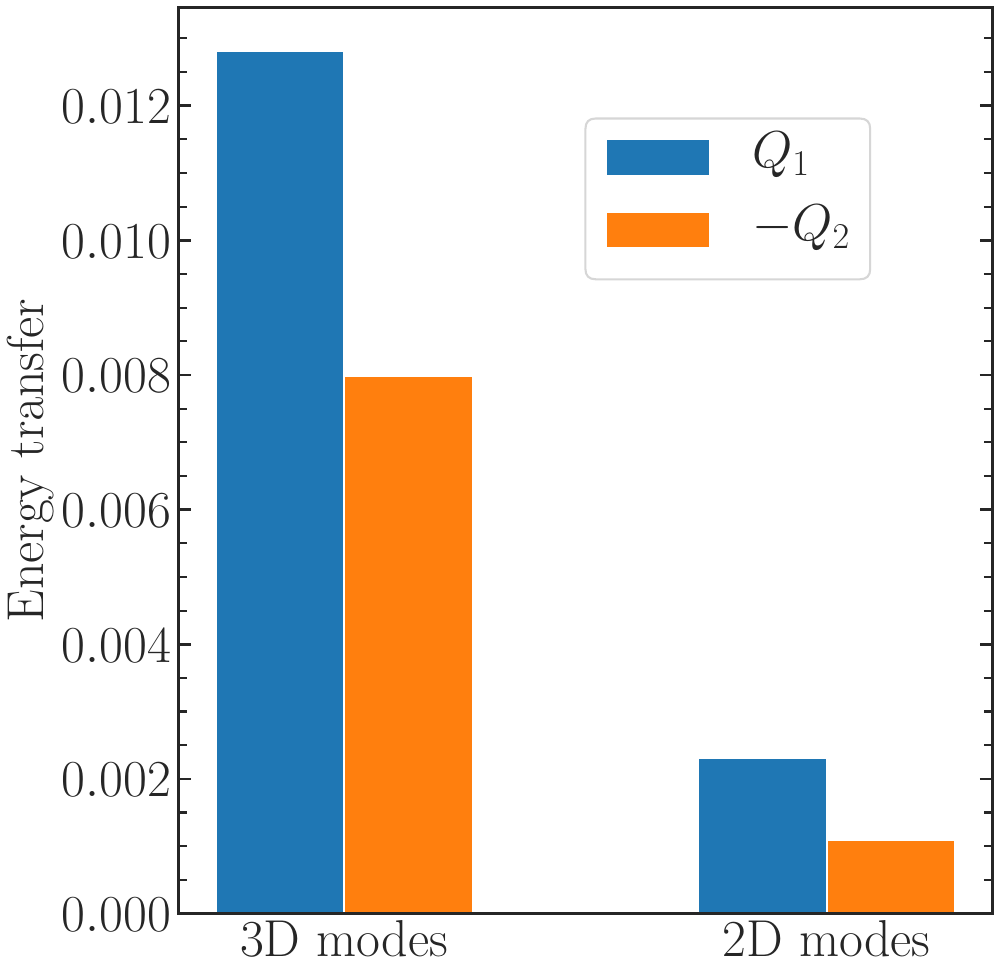}
    \caption{Energy transfer rates, classified by $2$D and $3$D modes, with further decomposition into the unstable and the stable modes. In the simulation with $Re\mathrm{=}5000$, time averaging is taken over $t\mathrm{=}200\textrm{--}708$. The stable-to-unstable mode fraction is larger for $3$D modes than for $2$D.}
    \label{fig:f8n}
\end{figure*}

The $3$D modes are more efficient than the $2$D modes in extracting energy from the mean flow, as Fig.~\ref{fig:f8n} shows.  Additionally, the stable-to-unstable mode fraction is higher for $3$D modes.  The presence of the energy reversal by the stable modes in $3$D confirms that the stable modes are not the manifestations of the traditional inverse cascade of $2$D turbulence; the former is a direct fluctuation-to-mean transfer of energy, whereas the latter is a fluctuation-to-fluctuation transfer that takes multiple iterations and is a nonlinear process. The quantities measured in Fig.~\ref{fig:f8n} are the rates of energy transfer directly between the mean and the fluctuations; we decompose such a linear energy transfer rate into the contributions of the unstable and stable modes, going beyond the simple traditional spectral transfer. The significant direct energy reversal in $3$D by the stable modes is a somewhat surprising and important finding of this paper.

\section{Nonlinear energy transfer to stable modes}
\label{sec:near-resonant}

Since the stable modes are found to contribute significantly to the energetics and transport, what nonlinear process transfers energy to the stable modes? Established\cite{verma2019, gome2023} diagnostics of energy transfer in wavenumber space fail here, because they only inform about the energy transfer to a given wavenumber, whereas, at any wavenumber, the total fluctuation is composed of many different eigenmodes, making it desirable to put an eigenmode filter in the transfer analysis.

\begin{figure*}
    \centering
    \includegraphics[width=0.95\textwidth]{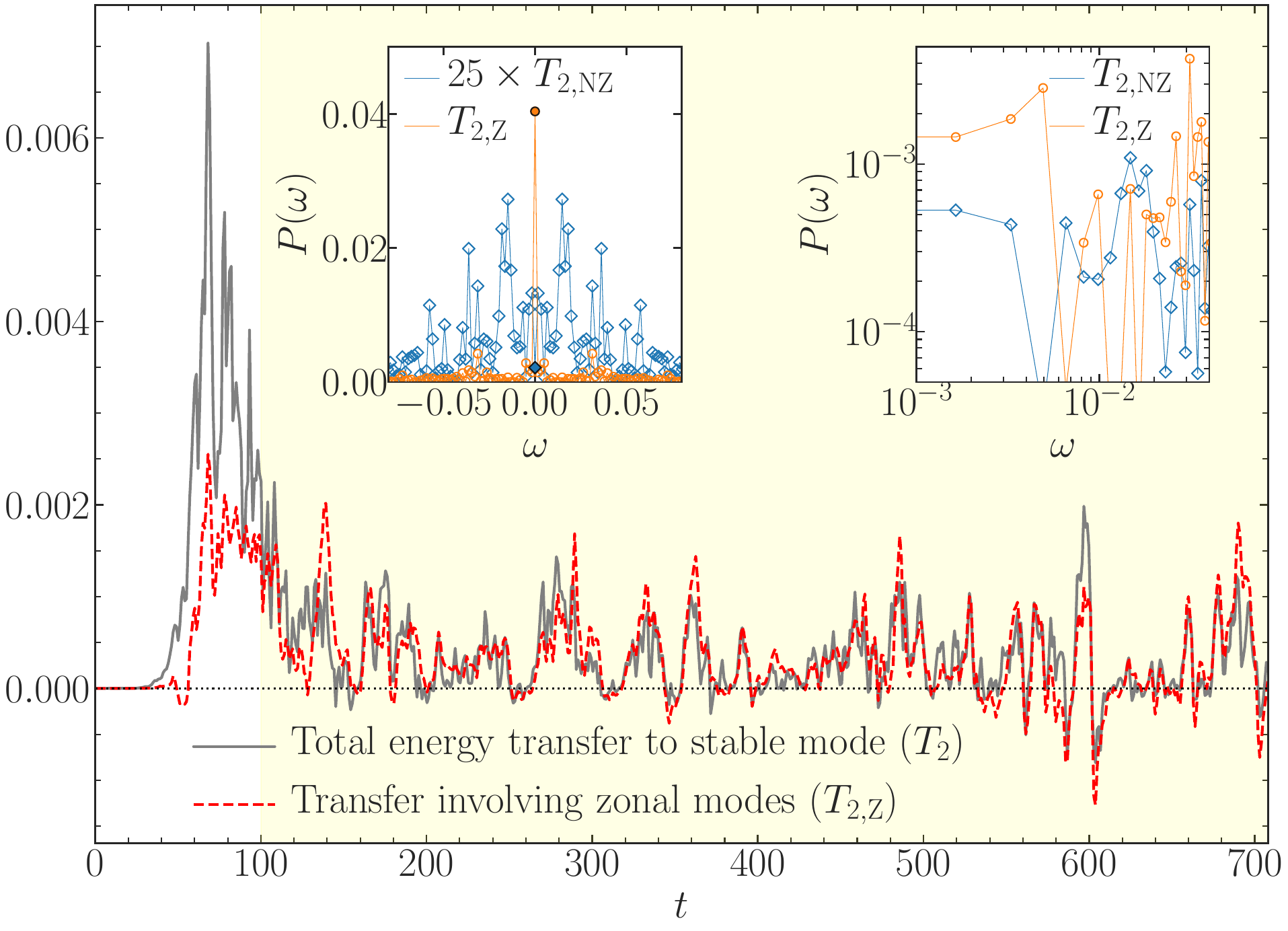}
    \caption{Nonlinear energy transfer $T_2$ to the stable mode, largely channeled via spanwise-only (zero-frequency) fluctuation $T_{2,\mathrm{Z}}$, which becomes dominant after $t\gtrsim 100$, shown in the yellow-shaded region. The coupling to zero-frequency modes, labeled zonal modes, is further probed using a power spectrum $P(\omega)$ of energy transfer in frequency $\omega$, shown in the insets, with linear scales (left) and with logarithmic scales (right), where $T_{2,\mathrm{NZ}} \mathrm{=} T_2\mathrm{-}T_{2,\mathrm{Z}}$ represents a purely non-zonal transfer. At $\omega\mathrm{=}0$, filled markers are used. Zonal-coupling dominates at $|\omega| \mathrm{\approx} 0$, implying that nonlinear interactions feeding energy to stable modes are primarily coherent.}
    \label{fig:f9}
\end{figure*}

The total nonlinear energy transfer $T_2$ from all the fluctuations to a stable mode $j\mathrm{=}2$ at $\mathbf{k} \mathrm{=} (0.2, 0)$ is shown in Fig.~\ref{fig:f9}, using the expressions in Eqs.~\eqref{eq:dtbetajenergy} and ~\eqref{eq:Tj}. The overlaid curve is the energy transfer rate $T_{2,\mathrm{Z}}$ based on triad interactions where one of the triad members is a fluctuation with $k_x\mathrm{=}0$, channeling energy to the stable mode. This is found by restricting the summation in Eq.~\eqref{eq:Tj} such that either $\mathbf{k}'\mathrm{=}(0, k_y')$ or $\mathbf{k}''\mathrm{=}(0, k_y'')$. We note that $k_x\mathrm{=}0$ has zero growth rate and zero frequency. This wavenumber, and, in particular, the first Fourier mode number along the $y$-axis, with $k_x\mathrm{=}0$, is observed to have a large $x$-component in the turbulent flow, and may be called a zonal mode, analog to secondary instabilities in fusion plasmas.\cite{rogers2000,pueschel2013,terry2018}  A difference, however, exists: here, these zonal modes contribute to the Reynolds stress, as opposed to those of fusion plasmas, which do not transport momentum and, hence, the contribution of the zonal modes to the stress is null.\cite{terry2021} The main similarity is that both of the systems have qualitatively similar anisotropic dispersion relation and both have zonal modes whose linear frequencies are zeros in the inviscid/ideal limit (when one considers small-scale dissipation coefficient, as well, then the complex frequencies feature damping rates).\cite{terry2018,terry2021}

We remark that the stable modes are excited in our system primarily via nonlinear interaction with the $k_x\mathrm{=}0$ modes, once the initial transient phase is complete by around $t\mathrm{=}100$ (Fig.~\ref{fig:f9}). Since the same triadic interaction continuously feeds energy to the stable mode in the fully saturated phase, one may infer that the process of nonlinear formation of the stable mode is rather coherent, and, to first order, does not involve changing wavevectors; the triad driving the stable mode is largely the same. Such an understanding of stable-mode formation is difficult to obtain in spectral-only representation, as at a given KH-unstable wavenumber, there also exists the stable mode, along with the continuum modes. This does not mean that it is uninsightful to perform traditional diagnostic of energy transfer in wavenumber space by integrating fluctuations along the direction of inhomogeneity.  In fact, some recent shear-flow studies have shown an interesting finding of energy circulation over angles on the wavevector plane, with the magnitude of the wavevector largely unchanged---a process that has been dubbed as ``transverse cascade." \cite{horton2010, mamatsashvili2014, mamatsashvili2016, gogichaishvili2017, gogichaishvili2018, mamatsashvili2020} One may investigate the connection between such a cascade and the stable modes to learn if the stable modes favor or impede the transverse cascade.  Such a study, however, is beyond the scope of the present paper.

The zero-frequency mode coupling leads to a near-resonant energy transfer to the stable mode,\cite{terry2021, terry2018, li2021, li2022} see insets of Fig.~\ref{fig:f9}, where both time-steady and slow time-evolution of energy transfers are dominantly zonal-coupled. Such a coupling allows a large amount of energy to be transferred coherently from the unstable mode. This finding is of striking consequence, as it has a direct potential to allow building a reduced, predictive model of saturation levels of unstable and stable modes, as has recently been achieved in the context of fusion microturbulence.\cite{terry2021, terry2018, li2021, pueschel2021, li2022, li2023}  The existence of zero-frequency mode coupling and near-resonant energy transfer in $3$D hydrodynamic turbulence, as found here, is surprising,\cite{biskamp1995} and may be considered a finding of the foremost importance.

\section{Vortex stretching, cascade, and viscous dissipation}
\label{sec:vortex-stretching}

After quantifying the effect of stable modes in energetics and transport, we now perform an ensemble of numerical experiments. 

We randomly select five snapshots in the turbulent phase of the standard simulation to construct an ensemble. In this ensemble, we project the state vector at each of those times onto the stable eigenmodes at wavenumbers $\mathbf{k}\mathrm{=}(0.2, 0)$ and $\mathbf{k}\mathrm{=}(0.2, \pm 0.2)$, as these wavenumbers dominate energetically (Fig.~\ref{fig:f7n}). Then, we delete these stable modes from the chosen state vector, and use the resulting modified state vector as an initial condition to start a new simulation.  The restarted simulations, therefore, produce responses to an impulsive zeroing of the stable-mode amplitudes. For a second ensemble, we repeat this process, projecting out this time the unstable modes from the same wavenumbers at the same five snapshots. In both ensembles, the five responses of each impulsive zeroing procedure are then averaged to produce two ensemble-averaged responses, one from zeroing the stable modes and another from zeroing the unstable modes.

\begin{figure*}
    \centering
    \includegraphics[width=0.7\textwidth]{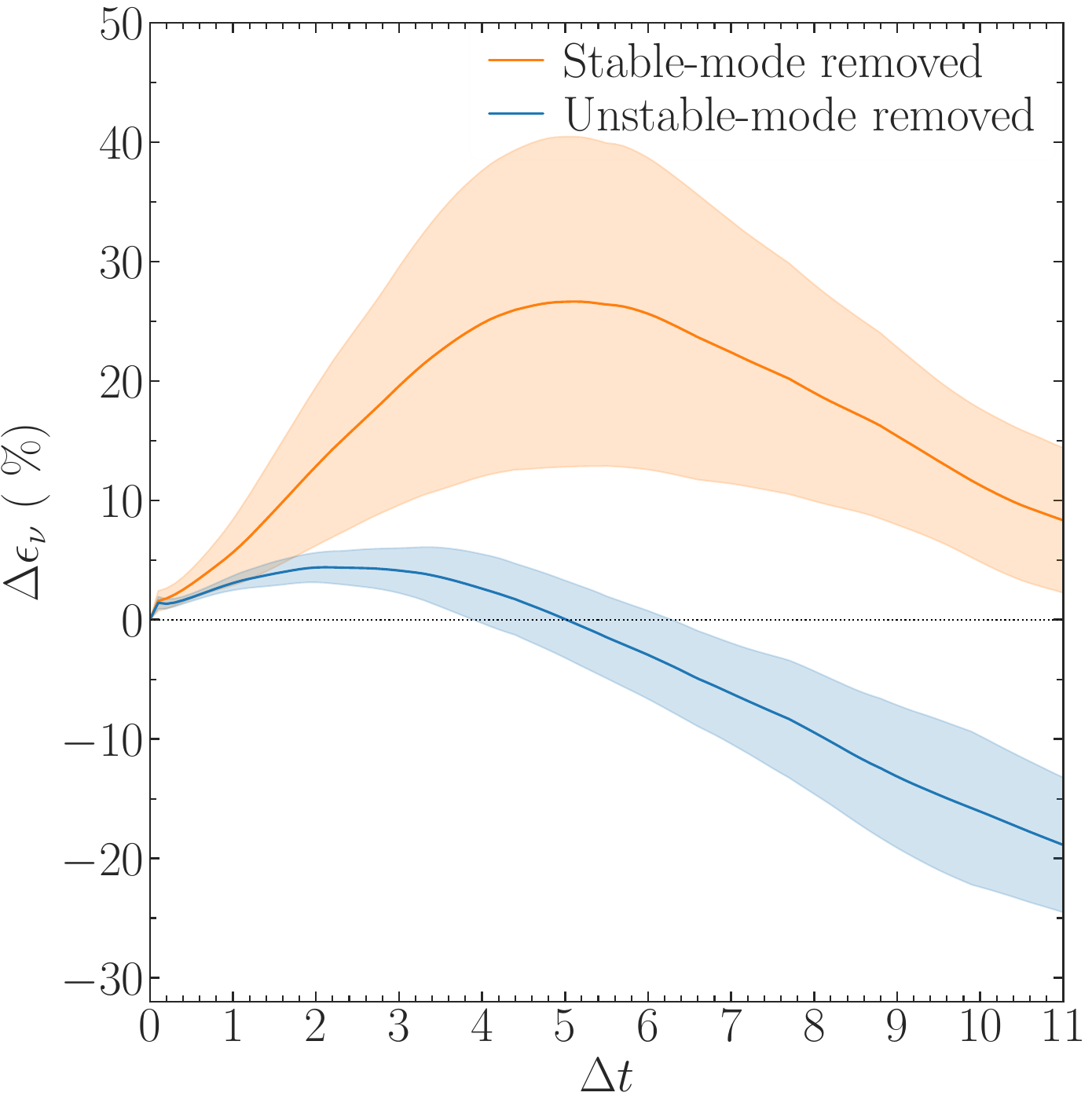}
    \caption{Percentage changes in small-scale viscous dissipation rates. The mean (solid curve) and one standard deviation (shaded) is found by simulating an ensemble of $10$ different $3$D simulations, where either stable or unstable modes are deleted instantaneously at a randomly selected time, and the simulation resumed to measure the effect on the nonlinear cascade via the small-scale dissipation rate. For all $\Delta t$ shown, stable-mode-removed simulations (orange) show higher dissipation rates than that in a standard simulation, corresponding to a positive $\Delta \epsilon_\nu$; lower dissipation rates are observed after an initial impulsive transient in unstable-mode-removed simulations (blue). All simulations were performed with $Re\mathrm{=}300$.}
    \label{fig:f10}
\end{figure*}

We then evaluate the viscous dissipation rate, computed at wavenumbers other than $(0.2, 0)$ and $(0.2, \pm 0.2)$ so that the initial dissipation rates in the mode-removed and standard simulations are identical.  The time-evolution of the dissipation rate $\Delta \epsilon_\nu$ is compared in Fig.~\ref{fig:f10} between the two ensembles. Larger small-scale dissipation rates are observed when stable modes are impulsively removed, similar to a recent finding in $2$D turbulence.\cite{tripathi2022a}  Although removing the unstable modes initially causes a minor positive change in dissipation rate, the real impact of the unstable-mode removal is quickly seen as a reduced dissipation rate---lower than in the standard simulation.  This result is straightforward to explain: first, when an eigenmode of significant amplitude is removed, the resulting large-scale flow structure for a brief moment nonlinearly strains the other scales of the flow. Second, \textit{nota bene}, the linear interaction of the unstable modes (energy source) or stable modes (energy sink) with the mean flow is impulsively impaired, which immediately results in either a depletion of fluctuation energy or a surplus. This second process dominates over the first one after $\Delta t \approx 3$. Soon after $\Delta t \approx 5$, we observe different behaviors in the dissipation rate---a set of curves lying in the positive $\Delta \epsilon_\nu$ region, and the other set lying in the negative. The negative is because the stable modes efficiently return the fluctuation energy to the mean flow; the positive change is due to the enhanced forward cascade of energy that is extracted from the mean flow by the unstable modes, which are not countered by the stable modes here. Around $\Delta t\mathrm{=} 5$, it is observed that the stable-mode removed simulations, on average, feature the highest dissipation rate, whereas the unstable-mode removed simulations, on average, show a change in the dissipation rate that is nearly null.

\begin{figure*}
    \centering
    \includegraphics[width=0.45\textwidth]{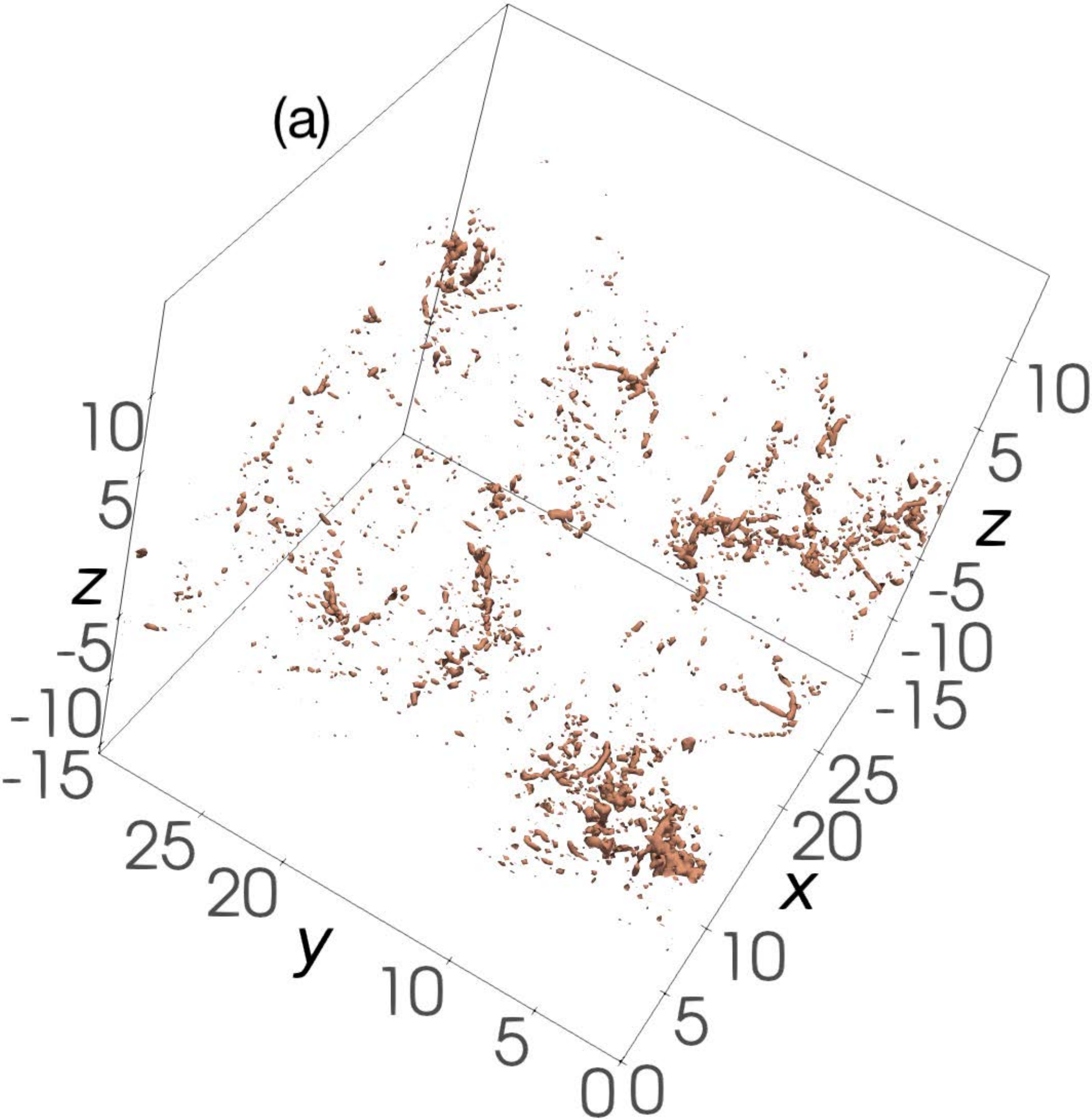}
    \includegraphics[width=0.45\textwidth]{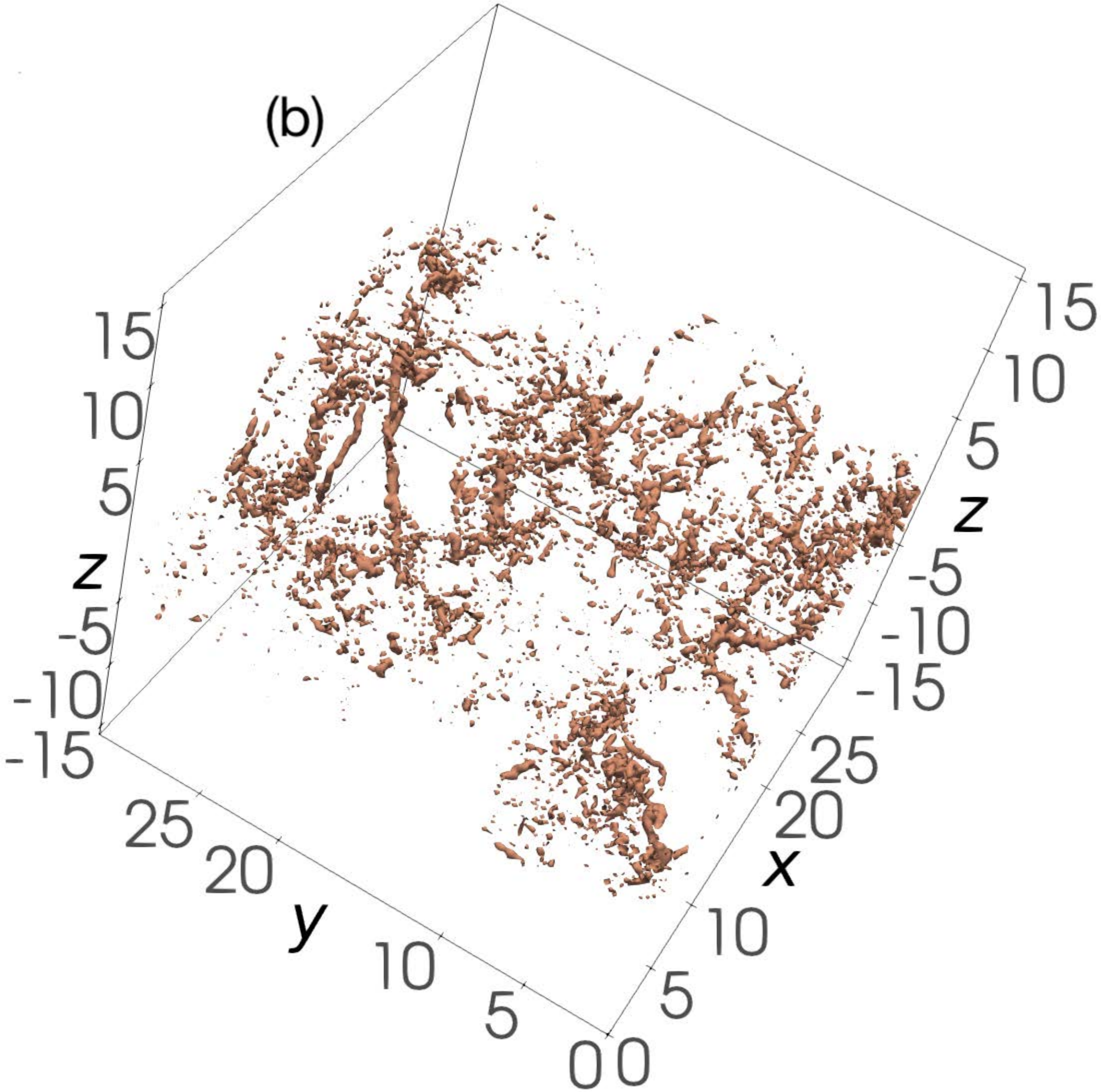}
    \includegraphics[width=0.45\textwidth]{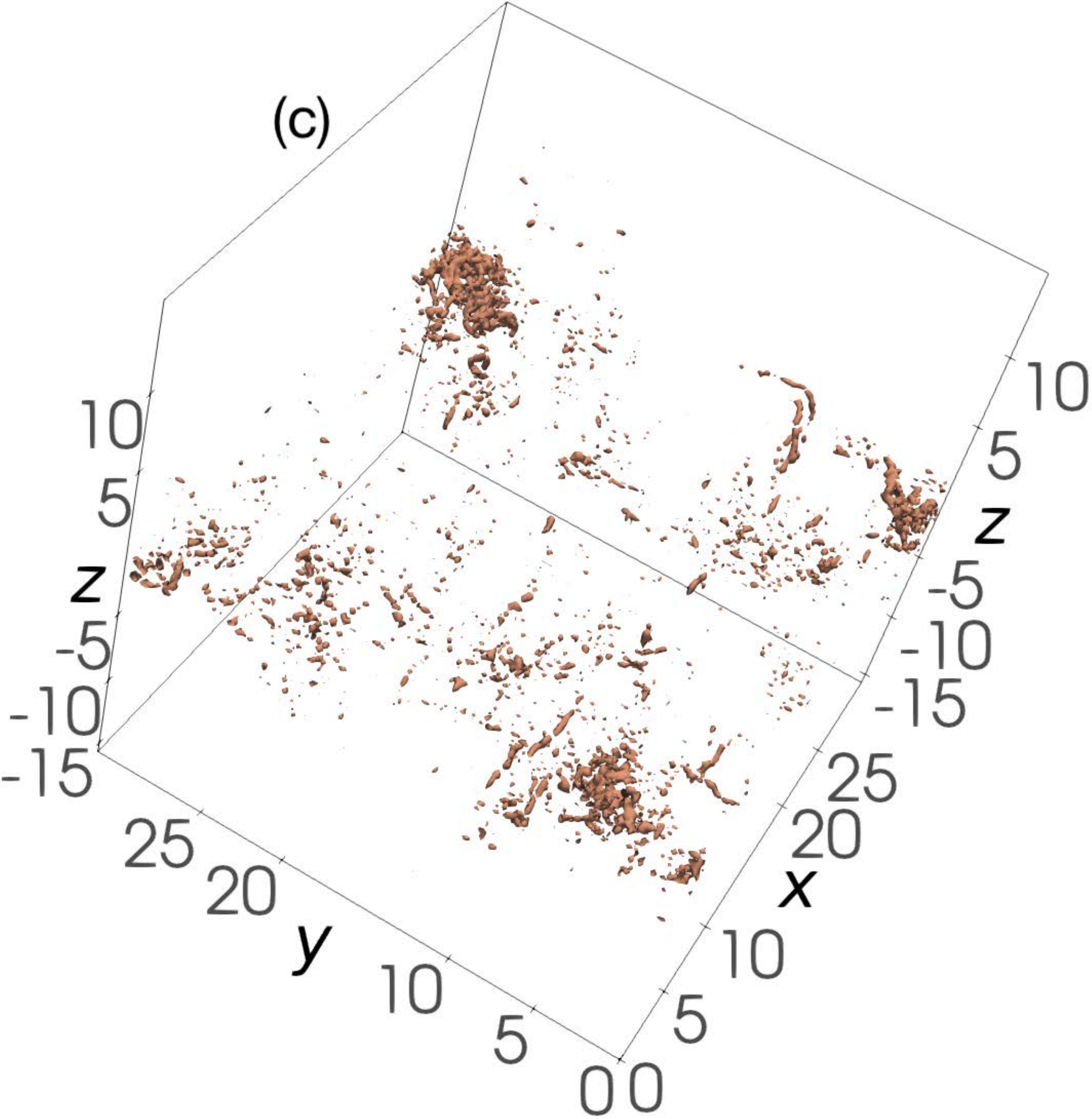}
    \caption{Isocontours of squared vorticity in simulations at a time when the viscous dissipation rate peaks. The same isocontour level is chosen in all panels. In the standard simulation in (a), vortex stretching is suppressed compared to a stable-mode-removed simulation in (b), where thin and elongated filamentary vortices are prominent.  In an unstable-mode-removed simulation in (c), the long filamentary structures are not as pronounced as in (b). The unstable modes, in the absence of stable modes, rapidly stretch and thin out the vortex tubes; in contrast, the stable modes, in the absence of unstable modes, deplete the fluctuation energy.}
    \label{fig:f11}
\end{figure*}

\begin{figure*}
    \centering
    \includegraphics[width=0.99\textwidth]{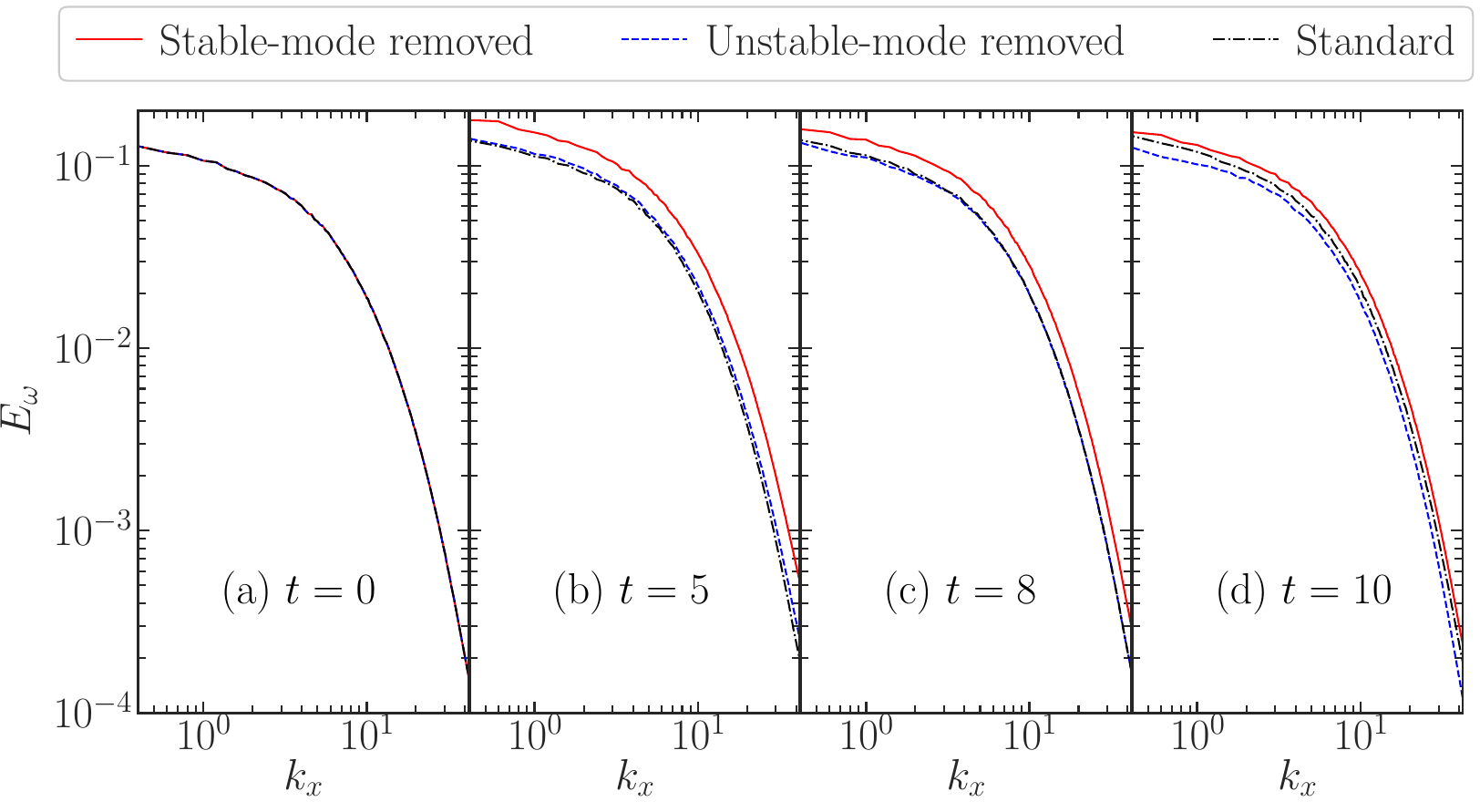}
    \caption{Time evolution of enstrophy $E_\omega = \langle |\hat{\mathbf{\omega}}|^2\rangle_{y,z}$ spectra in three simulation-continuations: stable-mode removed, unstable-mode removed, and standard.  (a) All simulations are restarted with an identical initial condition, except for the removed mode. The mentioned-modes are removed only from the wavenumbers $k_x=0.2$ and $k_y=\{0, \pm 0.2\}$. The spectra are, therefore, shown for $k_x>0.2$. (b)--(d) Impulsive responses to the mode-removal are most pronounced when the unstable modes are removed, leading to enhanced turbulence at all scales. The $k_y$-spectra of $(x,z)$-averaged enstrophy are similar (not shown).}
    \label{fig:f12}
\end{figure*}

After quantitative analysis of dissipation rates, we show visualizations of squared vorticity (whose volume integral is proportional to the dissipation rate) and compare the physical structures of turbulence across the simulated responses. At the time when the dissipation rate is maximal, the isocontours of squared vorticity show, in a stable-mode impulsively-zeroed simulation, an uninhibited stretching of vortex tubes by the unstable modes, compare Fig.~\ref{fig:f11}(b) with Fig.~\ref{fig:f11}(a). In the unstable-mode impulsively-zeroed simulation in Fig.~\ref{fig:f11}(c), a large-scale structure tends to develop instead of vortex stretching that forms prominent small scales. We note that the vortex-tube stretching is a purely $3$D phenomenon, and is associated with the forward cascade of energy in $3$D hydrodynamic turbulence.  A movie, found in the online supplemental material of this paper, shows several snapshots of stable-mode-removed simulations, where we observe prominent episodes of vortex stretching.

To quantify the changes in vortex dynamics at smaller scales in response to the large-scale-eigenmode removal, we measure enstrophy spectra in three simulation-continuations, and show a comparison of their time evolution in Fig.~\ref{fig:f12}. Although started with an identical initial condition, the simulation with stable-mode-removal rapidly evolves, as a nonlinear response to the mode-deletion, to feature increased level of enstrophy at all scales. The unstable-mode removed simulation also evolves; however, the increment in the small-scale activity, early on, is relatively small, and, at later times, the simulation, in fact, shows decreased enstrophy.

One may interpret the findings of the simulation experiments---increased filamentary vortex structures due to the unstable modes, and reduced filamentary vortex structures due to the stable modes---as the competing effects of forward transfer of energy from the mean flow to the fluctuation by the unstable mode and inverse transfer of energy by the stable mode in reversed direction.  The nonlinear process then responds to this linear physics.

\section{Discussion}
\label{sec:discussion}
Some broader implications and interpretations of the results of this work shall now be provided.

\subsection{Cascades, Couplings, and Competitions}
We have shown with various quantitative measures that only a fraction of unstable-mode energy is cascaded to small scales.  A majority, near or greater than $70\%$, of the instability-extracted energy from the mean flow to the large-scale fluctuations is, in reality, nonlinearly transferred to the conjugate-stable modes. This nonlinear mode-coupling at the instability-scale controls the rate of the small-scale energy cascade right at its inception. 

It remains true that larger Reynolds numbers push the Kolmogorov dissipation length scale to smaller scales, thus allowing more scales to be dynamically relevant in the turbulence.  The small-scale turbulence is indeed affected by the Reynolds number; however, the measures of interest, such as the energy injection rate in the cascade channel where it is formed and the momentum transport rate, are under the territorial jurisdiction of the fluctuations at the instability scale. At such a large scale, the conjugate-stable modes are excited, first via a parametric drive from the unstable modes through the mode-coupling coefficient $C_{211}(\mathbf{k}, \mathbf{k}')$ in the stable-mode evolution equation
\begin{equation}
\partial_t \beta_2(\mathbf{k}) = -\gamma (\mathbf{k}) \beta_2(\mathbf{k}) + \sum_{l}B_{2l}(\mathbf{k}) \beta_l(\mathbf{k}) + \sum_{\mathbf{k}', \mathbf{k}'': \mathbf{k}'+\mathbf{k}''=\mathbf{k}} C_{211}(\mathbf{k}, \mathbf{k}') \beta_1' \beta_1'' + \ldots,
\end{equation}
where $C_{211}(\mathbf{k}, \mathbf{k}')$ is independent of viscosity, and, therefore, favors excitation of stable modes. Whether this excitation becomes significant or not depends on the relative magnitudes of $\gamma$, $B_{2l}$, and $C_{211}(\mathbf{k}, \mathbf{k}')$.\cite{terry2006}  

At the large instability scale, with decreasing viscosity, $B_{2l}$ naturally becomes smaller; this is different from the singular limit of small-scale dissipation that does not vanish with decreasing viscosity. Therefore, the large-scale-stable-mode excitation is only weakly impacted once the viscosity is sufficiently low, as we have found in Fig.~\ref{fig:f3}. In addition, the small-scale dissipation rate, as well as the scales where intense viscous dissipation occurs, are both controlled by the energy extracted from the mean flow-gradient; the extraction rate, however, depends on the excitation levels of the large-scale stable modes, as shown in Figs.~\ref{fig:f10} and \ref{fig:f11}.  Thus, the Reynolds number, although it spawns more small scales, only asymptotically and weakly impacts the large-scale-stable-mode physics, as the stable modes are coupled to the unstable modes via a nonlinear mode-coupling coefficient whose origin lies in the overlap of the eigenmodes of an instability that is entirely inviscid. Although a competition exists between small-scale cascades and nonlinear coupling among the roots of the inviscid instability, the latter dominates over the former as long as the Reynolds number is not so small that the flow becomes near-laminar, strongly breaking the symmetry between the unstable and stable modes---this latter case does not appear in high-Reynolds-number turbulent flows in natural systems.

\subsection{Dimensionality, Density of states, Drives, and Dynamics} 
For a system with $n$ spatial dimensions (physical coordinate axes), we conjecture that the number of KH-unstable wavenumbers grows following an approximate power law $(k/k_0)^{n-1}$, where $k$ is the magnitude of the wavevector, and $k_0\mathrm{=}2\pi/L$ is the lowest wavenumber in the domain of size $L$.  The increment in the number of KH-unstable wavenumbers follows from the density-of-states effect, associated with each dimension (Fig.~\ref{fig:f2}); one of the axes is not counted because that axis represents the direction of inhomogeneity of the mean shear flow.

Such increased number of the KH-unstable wavenumbers put the mean shear-flow under additional stress, all demanding energy and momentum redistribution. Turbulence with increased spatial dimensions, hence, extracts more energy from the mean flow and transports more momentum across the shear layer---i.e., the turbulence features increased energy drive and dynamics. In decaying turbulence, the slowly growing wavenumbers of higher spatial dimension (greater than two) may not get an opportunity to attain high amplitudes in comparison to the $2$D-mode amplitudes.  However, for a sufficiently forced mean-flow, such a discriminatory scenario is precluded; as evidenced in Figs.~\ref{fig:f5n} and~\ref{fig:f8n}, the larger density of states of $3$D modes manifests in increased levels of momentum transport and energy transfer.

The dynamics of shear-flow turbulence can be substantially different depending on the linear process driving the turbulence. We shall remark here on two important processes: the exponential instability growth and the transient non-modal growth.  When the Kelvin-Helmholtz instability is present, as in this paper, the instability rapidly grows exponentially [Figs.~\ref{fig:f3}(a) and (c), blue curves]. Such a rapid exponential growth over longer times supersedes the here-inconsequential transient growth due to non-modal effects.\cite{arratia2013, kaminski2014}  The latter are crucial in modally stable non-normal flows,\cite{gustavsson1991, chagelishvili1997, chagelishvili2003} e.g., shear flows with no inflection point. With the fluctuations dominated by the unstable modes, they nonlinearly excite conjugate-stable modes, with nearly twice\cite{terry2006} the instability-growth rate [Figs.~\ref{fig:f3}(a) and (c), orange curves].  Such stable modes, which are excited by the nonlinearity, are fundamentally different from the linear, transient growth of fluctuations. We also note that the conjugate-stable modes here continuously transfer energy from the fluctuations to the mean flow, as seen in Fig.~\ref{fig:f6n}(c). This continuous process differs from the transient non-modal growth.

\section{Conclusions}
\label{sec:conclusions}
The ubiquitous Kelvin-Helmholtz instability is traditionally assumed to saturate nonlinearly by cascading to small scales the unstable-mode energy, in its entirety. Recently, a majority of the unstable-mode energy has been reported to be returned to the mean flow from the fluctuation by the linearly stable eigenmodes that are nonlinearly excited in two-dimensional turbulence.\cite{fraser2017,tripathi2022a, tripathi2023} Whether such a consequential process exists in $3$D fluid turbulence, as well, is the central question addressed in this work. Here, the linearly stable $3$D modes, excited to significant amplitudes, are shown to exhibit similarities with the reports of $2$D turbulence, as well as new signatures that are fundamentally $3$D. The energy transfer path to stable modes in $3$D is wholly different: in the turbulent phase, the transfer occurs via modes, we label as zonal flows, that vary only in the direction orthogonal to the $2$D shear flow.  Such perturbations have zero linear frequencies due to the anisotropic dispersion relation of the inviscid $3$D Kelvin-Helmholtz instability.  However, because of their zero frequencies, they allow near-resonant energy transfer from unstable to conjugate-stable modes, saturating the instability in $3$D.

The transfer of energy by the stable modes is directly from the instability-scale fluctuation to the mean flow-gradient, which may be interpreted as an \textit{inverse transfer of energy}.  This inversion is a linear process, different from the traditional nonlinear energy cascade to small scales. We have shown here not only the existence of such inverse transfer of energy in $3$D, but that they are more efficient for the $3$D perturbations than for the $2$D perturbations.

Vortex-tube stretching, an inherently $3$D process, is suppressed by the stable modes.  This has been demonstrated through a series of numerical experiments, where we impulsively zeroed either stable or unstable modes in a nonlinear simulation, at randomly selected times, and resumed the simulations to obtain statistical responses in the small-scale activity.  The viscous dissipation rates and spectral amplitudes of enstrophy increase with stable-mode removal, and decrease with unstable-mode removal. Further, $3$D visualizations of vortex dynamics revealed that, in the absence of the stable modes, the vortex tubes are significantly stretched and continuously thinned out; the stable modes counter this mechanism and remove the thin filamentary structures, attempting to form large-scale structures.

Transporting up-gradient momentum in $3$D, the stable modes counteract the instability-driven down-gradient momentum transport more effectively than they do in 2D ($\mathrm{\approx} 70\% $ vs. $\mathrm{\approx} 50\%$). This competition occurs continuously in time and can occasionally reverse the net transport direction, bearing important implications for shear-flow turbulence in geo- and astrophysical environments.

For future work, this study raises important questions regarding the impact of stable modes in other three-dimensional problems such as $3$D MHD shear flows, where, analogous to the vortex stretching, there exists magnetic field amplification, which, at small scales, can be inhibited by the stable modes, favoring large-scale magnetic field generation\cite{tobias2013}; $3$D stratified shear flows, where the stable modes may serve as a zookeeper, managing a zoo of instabilities\cite{caulfield2021, klaassen1991, liu2022}; the ideal, magneto-rotational instability (MRI),\cite{oishi2020} where it is possible that the transport measured in numerical simulations of MRI-driven turbulence (see, e.g., Ref.~\cite{pessah2006}) has an unaccounted contribution from the stable modes, and there is an opportunity to build accurate transport models for astrophysical problems, magnetized\cite{ji2023, blandford2022} or not.

\begin{acknowledgments}
We are grateful to Keaton~Burns, Evan~H.~Anders, Daniel~Lecoanet, and Michael~J.~Gerard for useful discussions. This material is based upon work funded by the Department of Energy (Grant No.~DE-SC0022257) through the NSF/DOE Partnership in Basic Plasma Science and Engineering. A.E.F. acknowledges support from NASA HTMS Grant No.~80NSSC20K1280, and from the George Ellery Hale Postdoctoral Fellowship in Solar, Stellar and Space Physics at the University of Colorado, Boulder.  The simulations used the ACCESS supercomputing resources under Allocation No.~TG-PHY130027. 

The data that support the findings of this study are available from the corresponding author upon reasonable request.
\end{acknowledgments}



\begin{thebibliography}{9}

\bibitem{alexakis2018}
A.~Alexakis and L.~Biferale, Cascades and transitions in turbulent flows, {\em Phys. Rep.} \textbf{767}, 1 (2018).

\bibitem{kolmogorov1941}
A.N.~Kolmogorov, The local structure of turbulence in incompressible viscous fluid for very large Reynolds numbers {\em Dokl. Akad. Nauk S.S.S.R.} \textbf{30}, 299 (1941).

\bibitem{frisch1995}
U.~Frisch, {\em Turbulence: The Legacy of A.~N.~Kolmogorov} (Cambridge University Press, Cambridge, England, 1995).

\bibitem{kraichnan1967}
R.H.~Kraichnan, Inertial Ranges in Two-Dimensional Turbulence, {\em Phys. Fluids} \textbf{10}, 1417 (1967).

\bibitem{leith1968}
C.E.~Leith, Diffusion Approximation for Two-Dimensional Turbulence, {\em Phys. Fluids} \textbf{11}, 671 (1968).

\bibitem{waleffe1993}
F.~Waleffe, Inertial transfers in the helical decomposition, {\em Phys. Fluids A} \textbf{5}, 677 (1993).

\bibitem{smith1999}
L.M.~Smith and F.~Waleffe, Transfer of Energy to 2D Large Scales in Forced, Rotating 3D Turbulence, {\em Phys. Fluids} \textbf{11}, 1608 (1999).

\bibitem{smith2002}
L.M.~Smith and F.~Waleffe, Generation of Slow, Large Scales in Forced, Rotating, Stratified Turbulence, {\em J. Fluid Mech.} \textbf{451}, 145 (2002).

\bibitem{riley2000}
J.J.~Riley and M.-P.~Lelong, Fluid motions in the presence of strong stable stratification, {\em Annu. Rev. Fluid Mech.} \textbf{32}, 613 (2000).

\bibitem{horton2010}
W.~Horton, J.-H.~Kim, G.D.~Chagelishvili, J.C.~Bowman, and J.G.~Lominadze, Angular redistribution of nonlinear perturbations: A universal feature of nonuniform flows,  {\em Phys. Rev. E} \textbf{81}, 066304 (2010).

\bibitem{mamatsashvili2014}
G.R.~Mamatsashvili, D.Z.~Gogichaishvili, G.D.~Chagelishvili, and W.~Horton, Nonlinear transverse cascade and two-dimensional magnetohydrodynamic subcritical turbulence in plane shear flows,  {\em Phys. Rev. E} \textbf{89}, 043101 (2014).

\bibitem{mamatsashvili2016}
G.~Khujadze, G.~Chagelishvili, S.~Dong, J.~Jim\'enez, and H.~Foysi, Dynamics of homogeneous shear turbulence: A key role of the nonlinear transverse cascade in the bypass concept,  {\em Phys. Rev. E} \textbf{94}, 023111 (2016).

\bibitem{gogichaishvili2017}
D.Z.~Gogichaishvili, G.R.~Mamatsashvili, W.~Horton, G.D.~Chagelishvili, and G.~Bodo, Nonlinear Transverse Cascade and Sustenance of MRI Turbulence in Keplerian Disks with an Azimuthal Magnetic Field, {\em \apj} \textbf{845}, 70 (2017).

\bibitem{gogichaishvili2018}
D.Z.~Gogichaishvili, G.R.~Mamatsashvili, W.~Horton, and G.D.~Chagelishvili, Active Modes and Dynamical Balances in MRI Turbulence of Keplerian Disks with a Net Vertical Magnetic Field, {\em \apj} \textbf{866}, 134 (2018).

\bibitem{mamatsashvili2020}
G.R.~Mamatsashvili, G.D.~Chagelishvili, M.E.~Pessah, F.~Stefani, and G.~Bodo, Zero Net Flux MRI Turbulence in Disks: Sustenance Scheme and Magnetic Prandtl Number Dependence,  {\em \apj} \textbf{904}, 47 (2020).

\bibitem{biskamp1995}
D.~Biskamp and A.~Zeiler, Nonlinear Instability Mechanism in 3D Collisional Drift-Wave Turbulence, {\em Phys. Rev. Lett.} \textbf{74}, 706 (1995).

\bibitem{ng1996}
C.~Ng and A.~Bhattacharjee, Interaction of Shear-Alfven Wave Packets: Implication for Weak Magnetohydrodynamic Turbulence in Astrophysical Plasmas, {\em  Astrophys. J.} \textbf{465}, 845 (1996).

\bibitem{terry2004}
P.W.~Terry, Inverse energy transfer by near-resonant interactions with a damped-wave spectrum, {\em Phys. Rev. Lett.} \textbf{93}, 235004 (2004).

\bibitem{du2023}
S.~Du, H.~Li, X.~Fu, and Z.~Gan, Anisotropic Energy Transfer and Conversion in Magnetized Compressible Turbulence, {\em  Astrophys. J.} \textbf{948}, 72 (2023).

\bibitem{rubio2014}
A.M.~Rubio, K.~Julien, E.~Knobloch, and J.B.~Weiss, Upscale energy transfer in three-dimensional rapidly rotating turbulent convection, {\em Phys. Rev. Lett.}
\textbf{112}, 144501 (2014).

\bibitem{guervilly2014}
C.~Guervilly, D.~Hughes, and C.A.~Jones, Large-scale vortices in rapidly rotating Rayleigh-Benard convection, {\em J. Fluid Mech.} \textbf{758}, 407 (2014).

\bibitem{fraser2017}A.E.~Fraser, P.W.~Terry, E.G.~Zweibel, and M.J.~Pueschel, Coupling of damped and growing modes in unstable shear flow, {\em Phys. Plasmas} \textbf{24}, 062304 (2017).

\bibitem{fraser2018}A.E.~Fraser, M.J.~Pueschel, P.W.~Terry, and E.G.~Zweibel, Role of stable modes in driven shear-flow turbulence, {\em Phys. Plasmas} \textbf{25}, 122303 (2018).

\bibitem{fraser2021}A.E.~Fraser, P.W.~Terry, E.G.~Zweibel, M.J.~Pueschel, and J.M.~Schroeder, The impact of magnetic fields on momentum transport and saturation of shear-flow instability by stable modes, {\em Phys. Plasmas} \textbf{28}, 022309 (2021).

\bibitem{tripathi2022a}B.~Tripathi, A.E.~Fraser, P.W.~Terry, E.G.~Zweibel, and M.J.~Pueschel, Mechanism for sequestering magnetic energy at large scales in shear-flow turbulence, {\em Phys. Plasmas} \textbf{29}, 070701 (2022).

\bibitem{tripathi2022b}B.~Tripathi, A.E.~Fraser, P.W.~Terry, E.G.~Zweibel, and M.J.~Pueschel, Near-cancellation of up- and down-gradient momentum transports in magnetized shear flow turbulence due to stable modes, {\em Phys. Plasmas} \textbf{29}, 092301 (2022).

\bibitem{tripathi2023}
B.~Tripathi, A.E.~Fraser, P.W.~Terry, E.G.~Zweibel, M.J.~Pueschel, and E.H.~Anders, Nonlinear mode coupling and energetics of driven magnetized shear-flow turbulence, {\em Phys. Plasmas} \textbf{30}, 072107 (2023).

\bibitem{terry2006}P.W.~Terry, D.A.~Baver, and S.~Gupta, Role of stable eigenmodes in saturated local plasma turbulence, {\em Phys. Plasmas} \textbf{13}, 022307 (2006).

\bibitem{terry2021}
P.W.~Terry, P.-Y.~Li, M.J.~Pueschel, and G.G.~Whelan, Threshold Heat-Flux Reduction by Near-Resonant Energy Transfer, {\em Phys. Rev. Lett.} \textbf{126}, 025004 (2021).

\bibitem{terry2018}P.W.~Terry, B.J.~Faber, C.C.~Hegna, V.V.~Mirnov, M.J.~Pueschel, and G.G.~Whelan, Saturation scalings of toroidal ion temperature gradient turbulence, {\em Phys. Plasmas} \textbf{25}, 012308 (2018).

\bibitem{makwana2011}K.D. Makwana, P.W. Terry, J.-H. Kim, and D.R. Hatch, Damped eigenmode saturation in plasma fluid turbulence, {\em Phys. Plasmas} \textbf{18}, 012302 (2011).

\bibitem{whelan2018}G.G. Whelan, M.J. Pueschel, and P.W. Terry, Nonlinear Electromagnetic Stabilization of Plasma Microturbulence, {\em \prl} \textbf{120}, 175002 (2018).

\bibitem{makwana2014}K.D. Makwana, P.W. Terry, M.J. Pueschel, and D.R. Hatch, Subdominant Modes in Zonal-Flow-Regulated Turbulence, {\em \prl} \textbf{112}, 095002 (2014).

\bibitem{hatch2011prl}
D.R.~Hatch, P.W.~Terry, F.~Jenko, F.~Merz, and W.M.~Nevins, Saturation of gyrokinetic turbulence through damped eigenmodes, {\em \prl} \textbf{106}, 115003 (2011).

\bibitem{hatch2011}
D.R.~Hatch, P.W.~Terry, F.~Jenko, F.~Merz, M.J.~Pueschel, W.M.~Nevins, and E.~Wang, Role of subdominant stable modes in plasma microturbulence, {\em Phys. Plasmas} \textbf{18}, 055706 (2011).

\bibitem{li2021}
P.-Y~Li, P.W.~Terry, G.G.~Whelan, and M.J.~Pueschel, Saturation physics of threshold heat-flux reduction, {\em Phys. Plasmas} \textbf{28}, 102507 (2021).

\bibitem{li2022}
P.-Y Li and P.W. Terry, Assessing physics of ion temperature gradient turbulence via hierarchical reduced-model representations, {\em Phys. Plasmas} \textbf{29}, 042301 (2022).

\bibitem{qian2020}
T.M.~Qian and M.E.~Mauel, Observation of weakly damped modes using high resolution measurement of turbulence in a dipole confined plasma, {\em Phys. Plasmas} \textbf{27}, 014501 (2020).

\bibitem{chandrashekhar1961}
S.~Chandrasekhar, {\em Hydrodynamic and Hydromagnetic Stability} (Clarendon Press, Oxford 1961).

\bibitem{drazin2004}
P.~Drazin and W.~Reid, {\em Hydrodynamic Stability} (Cambridge University Press, Cambridge, 2004).

\bibitem{buzz2018}
M.~Buzzicotti, H.~Aluie, L.~Biferale, and M.~Linkmann, Energy transfer in turbulence under rotation, {\em Phys. Rev. Fluid} \textbf{3}, 034802 (2018).

\bibitem{liu2022}
C.~Liu, A.~Kaminski, and W.~Smyth, The butterfly effect and the transition to turbulence in a stratified shear layer, {\em J. Fluid Mech.} \textbf{953}, A43 (2022).

\bibitem{klaassen1985}
G.P.~Klaassen and W.R.~Peltier, The onset of turbulence in finite-amplitude Kelvin-Helmholtz billows, {\em J. Fluid Mech.} \textbf{155}, 1 (1985).

\bibitem{klaassen1991}
G.P.~Klaassen and W.R.~Peltier, The influence of stratification on secondary instability in free shear layers, {\em J. Fluid Mech.} \textbf{227}, 71 (1991).

\bibitem{mashayek2012a}
A.~Mashayek and W.R.~Peltier, The `zoo' of secondary instabilities precursory to stratified shear flow transition. Part 1 Shear aligned convection, pairing, and braid instabilities, {\em J. Fluid Mech.} \textbf{708}, 5 (2012).

\bibitem{mashayek2012b}
A.~Mashayek and W.R.~Peltier, The `zoo' of secondary instabilities precursory to stratified shear flow transition. Part 2 The influence of stratification, {\em J. Fluid Mech.} \textbf{708}, 45 (2012).

\bibitem{salehipour2015}
H.~Salehipour, W.~Peltier, and A.~Mashayek, Turbulent diapycnal mixing in stratified shear flows: The influence of Prandtl number on mixing efficiency and transition at high Reynolds number, {\em J. Fluid Mech.} \textbf{773}, 178 (2015).

\bibitem{smith2021}
K.M.~Smith, C.P.~Caulfield, and J.R.~Taylor, Turbulence in forced stratified shear flows, {\em J. Fluid Mech.} \textbf{910}, A42 (2021).

\bibitem{marston2008}
J.B.~Marston, E.~Conover, and T.~Schneider, Statistics of an Unstable Barotropic Jet from a Cumulant Expansion, {\em J. Atmos. Sci.} \textbf{65}, 1955 (2008).

\bibitem{cope2020}
L.~Cope, P.~Garaud, and C.~Caulfield, The dynamics of stratified horizontal shear flows at low P\'eclet number, {\em J. Fluid Mech.} \textbf{903}, A1 (2020).

\bibitem{garaud2016}
P.~Garaud and L.~Kulenthirarajah, Turbulent Transport in a Strongly Stratified Forced Shear Layer with Thermal Diffusion, {\em \apj} \textbf{821}, 49 (2016).

\bibitem{lucas2017}
D.~Lucas, C.P.~Caulfield, and R.R.~Kerswell, Layer formation in horizontally forced stratified turbulence: connecting exact coherent structures to linear instabilities, {\em J. Fluid Mech} \textbf{832}, 409 (2017).

\bibitem{caulfield2020}
C.P.~Caulfield, Open questions in turbulent stratified mixing: Do we even know what we do not know?, {\em Phys. Rev. Fluids} \textbf{5}, 110518 (2020).

\bibitem{garaud2018}
P.~Garaud, Double-diffusive convection at low Prandtl number, {\em Annu. Rev. Fluid Mech.} \textbf{50}, 275 (2018).

\bibitem{fraser2023}
A.E.~Fraser, S.A.~Reifenstein, and P.~Garaud, Magnetized fingering convection in stars, arXiv (2023). arXiv:2302.11610.

\bibitem{masson2018}
A.~Masson and K.~Nykyri, Kelvin-Helmholtz Instability: Lessons Learned and Ways Forward, {\em Space Sci. Rev} \textbf{214}, 71 (2018).

\bibitem{lecoanet2016}
D.~Lecoanet, M.~McCourt, E.~Quataert, K.J.~Burns, G.M.~Vasil, J.S.~Oishi, B.P.~Brown, J.M.~Stone, and R.M.~O{'}Leary, A validated non-linear Kelvin-Helmholtz benchmark for numerical hydrodynamics, {\em Mon. Not. R. Astron. Soc.} \textbf{455}, 4274 (2016).

\bibitem{pueschel2011}
M.J.~Pueschel, F.~Jenko, D.~Told, and J.~B\"uchner, Gyrokinetic simulations of magnetic reconnection, {\em Phys. Plasmas} \textbf{18}, 112102 (2011).

\bibitem{pueschel2014}
M.J.~Pueschel, D.~Told, P.W.~Terry, F.~Jenko, E.G.~Zweibel, V.~Zhdankin, and H.~Lesch, Magnetic reconnection turbulence in strong guide fields: Basic properties and application to coronal heating, {\em Astrophys. J., Suppl. Ser.} \textbf{213}, 30 (2014).

\bibitem{burns2020}
K.J.~Burns, G.M.~Vasil, J.S.~Oishi, D.~Lecoanet, and B.P.~Brown, Dedalus: A flexible framework for numerical simulations with spectral methods, {\em Phys. Rev. Res.} \textbf{2}, 023068 (2020).

\bibitem{paul2019}
E.~Paul, I.~Abel, M.~Landreman, and W.~Dorland, An adjoint method for neoclassical stellarator optimization, {\em J. Plasma Phys.} \textbf{85}, 795850501 (2019).

\bibitem{terry2009}P.W.~Terry, D.A.~Baver, and D.R.~Hatch, Reduction of inward momentum flux by damped eigenmodes, {\em Phys. Plasmas} \textbf{16}, 122305 (2009).

\bibitem{fuller2019}J.~Fuller, A.L.~Piro, and A.S.~Jermyn, Slowing the spins of stellar cores, {\em Mon. Not. R. Astron. Soc.} \textbf{485}, 3661 (2019).

\bibitem{pessah2006}M.E.~Pessah, C.-K.~Chan, and D.~Psaltis, The signature of the magnetorotational instability in the Reynolds and Maxwell stress tensors in accretion discs, {\em Mon. Not. R. Astron. Soc.} \textbf{372}, 183 (2006).

\bibitem{goodman1994}
J.~Goodman and G.~Xu, Parasitic Instabilities in Magnetized, Differentially Rotating Disks, {\em \apj} \textbf{432}, 213 (1994).



\bibitem{barker2019}
A.J.~Barker, C.A.~Jones, and S.M.~Tobias, Angular momentum transport by the GSF instability: non-linear simulations at the equator, {\em Mon. Not. R. Astron. Soc.} \textbf{487}, 1777 (2019).

\bibitem{verma2019}
M.K.~Verma, {\em Energy Transfers in Fluid Flows: Multiscale and Spectral Perspectives}  (Cambridge University Press, Cambridge 2019).

\bibitem{gome2023}
S.~Gom\'e, L.~Tuckerman, and D.~Barkley, Patterns in transitional shear turbulence. Part 1. Energy transfer and mean-flow interaction, {\em J. Fluid Mech.} \textbf{964}, A16 (2023).

\bibitem{rogers2000}
B.N.~Rogers, W.~Dorland, and M.~Kotschenreuther, Generation and Stability of Zonal Flows in Ion-Temperature-Gradient Mode Turbulence,  {\em \prl} \textbf{85}, 5336 (2000).

\bibitem{pueschel2013}
M.J.~Pueschel, T.~G$\ddot{\mathrm{o}}$rler, F.~Jenko, D.R.~Hatch, and A.J.~Cianciara, On secondary and tertiary instability in electromagnetic plasma microturbulence, {\em Phys. Plasmas} \textbf{20}, 102308 (2013). 

\bibitem{pueschel2021}
M.J.~Pueschel, P.-Y.~Li, and P.W.~Terry, Predicting the critical gradient of ITG turbulence in fusion plasmas, {\em Nucl. Fusion} \textbf{61}, 054003 (2021).

\bibitem{li2023}
P.-Y.~Li, M.J.~Pueschel, P.W.~Terry, and G.G~Whelan, On the role of mode resonances in regulating zonal-flow-moderated plasma microturbulence, {\em Nucl. Fusion} \textbf{63}, 026028 (2023).

\bibitem{arratia2013}
C.~Arratia, C.P.~Caulfield, and J.M.~Chomaz, Transient perturbation growth in time-dependent mixing layers, {\em J. Fluid Mech.} \textbf{717}, 90 (2013).

\bibitem{kaminski2014}
A.K.~Kaminski, C.P.~Caulfield, and J.R.~Taylor, Transient growth in strongly stratified shear layers, {\em J. Fluid Mech.} \textbf{758}, R4 (2014).

\bibitem{gustavsson1991}
L.H.~Gustavsson,  Energy growth of three-dimensional disturbances in plane Poiseuille flow, {\em J. Fluid Mech.} \textbf{224}, 241 (1991).

\bibitem{chagelishvili1997}
G.D.~Chagelishvili, A.G.~Tevzadze, G.~Bodo, and S.S.~Moiseev. Linear mechanism of wave emergence from vortices in smooth shear flows, {\em \prl} \textbf{79}, 17 (1997).

\bibitem{chagelishvili2003}
G.D.~Chagelishvili, J-P.~Zahn, A.G.~Tevzadze, and J.G.~Lominadze, On hydrodynamic shear turbulence in Keplerian disks: Via transient growth to bypass transition, {\em Astron. Astrophys.} \textbf{402}, 401 (2003).

\bibitem{tobias2013}
S.M.~Tobias and F.~Cattaneo, Shear-driven dynamo waves at high magnetic Reynolds number, {\em Nature} \textbf{497}, 463 (2013).

\bibitem{caulfield2021}
C.P.~Caulfield, Layering, Instabilities, and Mixing in Turbulent Stratified Flows, {\em Annu. Rev. Fluid Mech.} \textbf{53}, 113 (2021).

\bibitem{oishi2020}
J.S.~Oishi, G.M.~Vasil, M.~Baxter, A.~Swan, K.J.~Burns, D.~Lecoanet, and B.P.~Benjamin, The magnetorotational instability prefers three dimensions, {\em Proc. R. Soc. A.} \textbf{476}, 20190622 (2020).


\bibitem{ji2023}
S.~Ji, J.~Fuller, and D.~Lecoanet, Magnetohydrodynamic simulations of the Tayler instability in rotating stellar interiors, {\em Mon. Not. R. Astron. Soc.} \textbf{521}, 5372 (2022).


\bibitem{blandford2022}
R.~Blandford and N.~Globus, Ergomagnetosphere, ejection disc, magnetopause in M87 - I. Global flow of mass, angular momentum, energy, and current, {\em Mon. Not. R. Astron. Soc.} \textbf{514}, 5141 (2022).

\end{thebibliography}
\end{document}